\documentclass[fleqn,usenatbib]{mnras}

\usepackage{newtxtext,newtxmath}

\usepackage[T1]{fontenc}

\DeclareRobustCommand{\VAN}[3]{#2}
\let\VANthebibliography\thebibliography
\def\thebibliography{\DeclareRobustCommand{\VAN}[3]{##3}\VANthebibliography}

\usepackage{graphicx}	
\usepackage{amsmath}	
\usepackage{xcolor}
\usepackage{subfigure}
\usepackage{physics}
\usepackage[normalem]{ulem}

\definecolor{crimson}{HTML}{DC143C}

\definecolor{purple_cust}{HTML}{800080}

\usepackage{comment}
\usepackage{orcidlink}

\title[Nd I–V Atomic Data for Kilonovae]{Excitation and Recombination Data for Nd I–V with Applications to Kilonovae}

\author[N. Ferguson et al.]{
N. Ferguson$^{1}$\thanks{E-mail: niamh.ferguson@strath.ac.uk}\orcidlink{0009-0005-3148-513X},
L. P. Mulholland$^{2}$\orcidlink{0009-0003-2668-5589},
M. McCann$^{2}$\orcidlink{0000-0002-1532-1240},
C. P. Ballance$^{2}$\orcidlink{0000-0003-1693-1793}
and M. G. O'Mullane$^{1}$\orcidlink{0000-0002-2160-4546}
\\
$^{1}$Department of Physics, University of Strathclyde, 107 Rottenrow, G4 0NG, United Kingdom\\
$^{2}$Astrophysics Research Centre, School of Mathematics and Physics, Queen's University Belfast, Belfast BT7 1NN, Northern Ireland}

\date{Accepted XXX. Received YYY; in original form ZZZ}

\pubyear{\the\year{}}

\begin{document}
\label{firstpage}
\pagerange{\pageref{firstpage}--\pageref{lastpage}}
\maketitle

\begin{abstract}
Neodymium (Nd), a key r-process element, significantly influences the opacity and emergent spectra of kilonovae resulting from neutron star mergers. 
We present new atomic data for the first five ionization stages of Neodymium, with emphasis on recombination rate coefficients and electron-impact excitation relevant to non-local thermodynamic equilibrium (non-LTE) modelling. 
Using the relativistic atomic structure code {\sc autostructure}, we compute energy levels, radiative and dielectronic recombination rates using the isolated-resonance approximation, and excitation data for these ions.
Excitation datasets are constructed using both direct (distorted-wave) and resonant excitation approaches, allowing a systematic assessment of the impact of resonance contributions on effective collision strengths. For the case of Nd {\sc ii}, we compare collisional strengths with those produced by the Dirac Atomic R-matrix Codes ({\sc darc}).
These comparisons demonstrate the importance of resonance-mediated excitation, especially for low-charge ions where near-threshold resonances significantly enhance rates. 
In terms of recombination for higher ionization stages, dielectronic recombination (DR) is found to dominate over a wide temperature range, while radiative recombination (RR) is comparable with DR at low temperatures for near-neutral species.
Our results highlight the pivotal role of non-LTE atomic physics in interpreting kilonovae observations and constraining nucleosynthesis yields. 
This work strengthens the atomic foundation necessary for linking observed electromagnetic signals to the physics of compact object mergers and heavy element formation.

\end{abstract}




\section{Introduction}

The electromagnetic transient produced by a compact binary neutron star merger has generated interest as a site for rapid neutron capture to occur.
r-Process nucleosynthesis will occur in environments with a high density of free neutrons and sufficiently high temperature to synthesise elements heavier than iron (Z=26).
Kilonovae exhibit the key characteristics for an r-process production environment in which many heavy elements, such as lanthanides, are created.

The detection of the 2017 neutron star merger came from the gravitational wave observation GW170817, where the corresponding electromagnetic signal AT2017gfo, was observed by many telescopes \citep{smartt_kilonova_2017,pian_spectroscopic_2017}.
The heavy elements synthesised generate the transient across the optical, ultraviolet and near-infrared wavelengths \citep{metzger_electromagnetic_2010}.
Great effort has been made to identify the features of AT2017gfo and the model spectra from an hour post-merger \citep{kasen_origin_2017, banerjee_simulations_2020,banerjee_opacity_2022, banerjee_diversity_2024,sneppen_emergence_2024}.
The interpretation of kilonova spectra requires extensive atomic data for heavy r-process elements. 
Consequently, significant effort has been devoted to calculating atomic structures, opacities and transition probabilities of lanthanides for radiative transfer modelling \citep{tanaka_radiative_2013,tanaka_systematic_2020,nahar_theoretical_2024,Flors2026}.
However, few elements have been spectrally identified with only some spectral features have candidate ions proposed.
Elements such as Sr \citep{watson2019identification,domoto_signatures_2021,gillanders_modelling_2022,tarumi_non-lte_2023}, La \citep{domoto_lanthanide_2022, gillanders_modelling_2024}, Ce \citep{domoto_lanthanide_2022,gillanders_modelling_2024} and Y \citep{sneppen_discovery_2023} have been strongly proposed to give the features found in the early-time spectra, whilst He \citep{tarumi_non-lte_2023, perego_production_2022}, Cs \citep{smartt_kilonova_2017}, Te \citep{smartt_kilonova_2017} and Rb \citep{pognan_nlte_2023} have been proposed as other possibilities at this time. In the nebular phase, elements such as
Te \citep{hotokezaka_tellurium_2023, gillanders_modelling_2024,mulholland2026electron}, Se and W \citep{hotokezaka_tungsten_2022} have been proposed as an emission feature in the late-time spectra. 
There has been additionally some investigation of lanthanide impact at this time, with limited existing data \citep{Pognan_Kawaguchi_Wanajo_Fujibayashi_Jerkstrand_2025} and exploration of the light r-process signatures \citep{Jerkstrand_Pognan_Banerjee_Sterling_Grumer_Ferguson_Butler_Gillanders_Smartt_Kawaguchi_et_al._2026}.

Alongside the gamma-ray burst event GRB230307A, the kilonova AT2023vfi was also spectroscopically observed by JWST.
The JWST spectra was obtained at 29 and 61 days post merger providing a good insight into the non-LTE effects on spectra.
\cite{gillanders_analysis_2025} discusses possible candidate ions for emissions seen at 2.0218, 2.1874, and 4.4168 $\mu$m with a candidate list including light r-process ions.
Modelling of the nebular phase requires detailed recombination rates and effort has began in generating such data for various ion stages of the heavy elements \citep{banerjee_nebular_2025,singh_dielectronic_2025} with previous calculations already available such as Se \citep{Sterling_Witthoeft_2011}.
However, atomic data for ions relevant to kilonova modelling is still often limited and incomplete.
Photospheric investigations have already emphasised the importance of lanthanides in relation to the opacity and their enrichment of the opacity \citep{kasen_opacities_2013,tanaka_systematic_2020,barnes_kilonovae_2021}.
Due to this, one would expect their contribution to the nebular phase to be just as impactful especially in a lanthanide-rich ejecta.
The opacity for Nd {\sc i}-{\sc iv} was initially computed by \cite{kasen_opacities_2013} in the LTE approximation where opacity is the dominant process to include within the model.
Since then, further detail has been included in the opacity computations with lanthanide ions being calculated using other atomic codes such as HFR, FAC and others \citep{gaigalas_extended_2019,fontes_line-binned_2020,banerjee_opacity_2022,flors_opacities_2023,banerjee_diversity_2024,deprince_kilonova_2025}

Although many other elements have been proposed, there is scope for debate on the importance of Neodymium (Nd) throughout the kilonova lifetime and exploration of the role of Nd still remains current \citep{Fontes_Vieira_Fryer_Kathirgamaraju_Korobkin_Ristic_Wollaeger_2026}.

It remains to be seen whether the significance of Nd translates into the non-LTE phase and the emergent spectra.
Work by \cite{maison_calculation_2024} discusses the possibility of observing Nd {\sc iii} lines in the infrared, produced from the forbidden transitions of the ground configuration.
\cite{gillanders_constraints_2021} also expressed the importance of the M1 and E2 transitions in elemental identification and the need to include them within kilonovae modelling.
More recently, the work by \cite{Pognan_Kawaguchi_Wanajo_Fujibayashi_Jerkstrand_2025} discusses the details of the contribution that Nd {\sc ii} gives to spectral formation. The role of the lanthanide ions may produce many blended emission lines in the infrared region with similar strengths to the strong single line emission features produced by M1 transitions attributable to other r-process ions.

Here, atomic data for Nd ions {\sc i}-{\sc v} are presented in support of the identifications and modelling of the nebular phase.
Data is presented for recombination (dielectronic and radiative) computed using {\sc autostructure} \citep{badnell_breitpauli_2011} and excitation using three different models, including direct excitation and resonant excitation computed using {\sc autostructure} and excitation from the R-matrix method.
Details of the atomic structure are shown in section \ref{sec:structure}, recombination data is presented in section \ref{sec:RecResults} and excitation data presented in \ref{sec:ExcResults} followed by discussion on the collisional radiative modelling implications in section \ref{sec:radmodel} and general conclusions.
Comparisons between the excitation methods are discussed in section \ref{sec:ExcResults} with an R-matrix comparison drawn for Nd II in section \ref{sec:ndIIexc}. 
Finally, we present conclusions and an outlook for future work.

 \section{Atomic Structure}\label{sec:structure}
Lanthanide ions are characterised by their varied completion of the $4f$-shell.
The open f-shell brings complexity and sensitivity which makes obtaining the correct positioning of energy levels difficult.
Lanthanides have a large density of energy levels resulting in many resonances.
The $4f$ electrons are largely shielded by the outer $n=5$ orbitals ($5s$ and $5p$), resulting in relatively weak interaction with the external environment. 
However, the comparable magnitudes of electron–electron interactions and spin–orbit coupling in these ions mean that a pure LS coupling scheme is insufficient, necessitating the use of an intermediate coupling approach when computing the atomic structure.
The lanthanide elements have a large density of energy levels within their structure which can be difficult to depict accurately in theoretical calculations.
However, radiative transfer simulations require accurate atomic data to explore spectral features and so a methodology of computing data for lanthanide ions is essential for the application to kilonovae.

\subsection{{\sc autostructure} : Atomic structure}
The accuracy of atomic energy levels important when calculating atomic processes such as recombination.
Several methods exist to refine the atomic structure beyond that obtained from a default calculation, with the choice of approach strongly influenced by the availability of experimental data, such as measured energy levels.

By default, {\sc autostructure} uses a Thomas-Fermi-Dirac-Amaldi (TFDA) potential in the computation of each radial function.
The TFDA potential contains $\lambda$ scaling parameters on each orbital that by default is set to unity.
The orbital-specific $\lambda$ scaling parameters are introduced to modify the radial dependence of the model potential.
These parameters adjust the degree of screening of the nuclear charge that is experienced by each orbital which results in the radial function expanding/contracting.
The scaling parameters are treated as variational quantities and are optimized to reproduce observed or target energy levels, therefore improving the accuracy of the calculated wavefunctions and derived atomic data such as radiative and autoionization rates.

The internal {\sc autostructure} optimization scheme is based on the variational principle.
In the case that no external validation via experimental energy levels or other theoretical calculations are available, this optimization scheme is the best method for {\sc autostructure} and has been reliable for all ion stages calculated.
In {\sc autostructure}, the optimization scheme varies orbital scaling parameters ($\lambda$) from their unity default to minimize a target functional, typically the total or average energy of atomic configurations. 
Users can specify which $\lambda$ values to vary, if not all, and the code will evaluate the functional by generating orbitals, solving the atomic structure, and computing energies. 
This scheme uses the method of \cite{Powell_1965} or conjugate gradient algorithms to iteratively update $\lambda$ values until convergence is reached. 
The final optimized parameters yield more accurate wavefunctions and better agreement with experimental data. 
This process enhances the quality of computed energies, transition rates, and other atomic properties, making it essential for precise structure and spectroscopy calculations.

Obtaining a good atomic structure for the Nd ions resulted in some challenges.
The configurations used within the structure and the subsequent atomic processes calculations are as listed in table \ref{tab:configs}.
\begin{table}
	\centering
	\caption{Configurations used for each ion stage in the {\sc autostructure} calculations, and the configurations used in the {\sc{grasp}} structure for use within the R-matrix excitation calculation for Nd {\sc ii}. Ground configuration in bold.}
	\label{tab:configs}
	\begin{tabular}{lccr} 
		\hline
		Ion & Configurations & No. of Levels\\
		\hline
		Nd I & $\mathbf{4f^4 6s^2}, 4f^45d6s,4f^46s6p ,4f^35d6s^2$ & 3670\\
		Nd II & $\mathbf{4f^46s},4f^45d,4f^46p,4f^35d^2,4f^35d6s, $ & 6888\\
        &$4f^35d6p, 4f^36s6p$\\
        Nd II ({\sc grasp}) & $\mathbf{4f^46s},4f^45d,4f^35d^2,4f^36s^2,4f^35d6s,$ & 7518 \\
        & $4f^46p, 4f^35d6p,4f^36s6p,4f^36p^2$ & \\
		Nd III & $\mathbf{4f^4}, 4f^35d, 4f^36s, 4f^36p$ & 817\\
        Nd IV & $\mathbf{4f^3},4f^25d, 4f^26s, 4f^26p$ & 241\\
        Nd V & $\mathbf{4f^2}, 4f5d, 4f6s, 4f6p, 4f6d$ & 69\\
		\hline
	\end{tabular}
\end{table}

\begin{table*}
\centering
\caption{Comparison of calculated energy levels of Nd {\sc i} from AS in Ry with those available in NIST (Ry) }
\renewcommand{\arraystretch}{1.2}
\label{Ndi_levels}
\begin{tabular}{lcccccc}
\hline\hline
Configuration& Term& J&	$E_{calc}$ &	$E_{exp}$ &	$\Delta $ E  \\
\hline
$4f^46s^2$ & $^5I$ & 4 & 0.000 & 0.000 &  0 \\
           & $^5I$ & 5 & 0.01395034 & 0.01027960 &  0.00367074 \\
           & $^5I$ & 6 & 0.02935673 & 0.02156602 &  0.00779071 \\
           & $^5I$ & 7 & 0.04576265 & 0.03355008 &  0.01221257 \\
           & $^5I$ & 8 & 0.06281480 & 0.04600625 &  0.01680855 \\
$4f^35d6s^2$ & $^5L^o$ & 6 & 0.06338602 & 0.06164003 &  0.00174599 \\
           & $^5K^o$ & 5 & 0.06498020 & 0.06245819 &  0.00252201 \\
$4f^45d6s$ & $^7L$ & 5 & 0.06554465 & 0.07723312 &  0.01168847 \\
           & $^7L$ & 6 & 0.07411860 & 0.08306283 &  0.00894423 \\
$4f^35d6s^2$ & $^5L^o$ & 7 & 0.08024072 & 0.07656910 &  0.00367162 \\

\hline\hline
\end{tabular}
\end{table*}

\begin{table*}
\centering
\caption{Comparison of calculated energy levels of Nd {\sc ii} from AS in Ry with those available in NIST (Ry) }
\renewcommand{\arraystretch}{1.2}
\label{Ndii_levels}
\begin{tabular}{lcccccc}
\hline\hline
Configuration& Term& J&	$E_{calc}$ &	$E_{exp}$ &	$\Delta $ E  \\
\hline
$4f^46s$ & $^6I$ & 7/2 & 0.000 & 0.000 &  0 \\
         & $^6I$ & 9/2 & 0.00499713 & 0.00467781 &  0.00031932 \\
         & $^6I$ & 11/2 & 0.01455098 & 0.01339665 &  0.00115433 \\
         & $^4I$ & 9/2 & 0.01613847 & 0.01503777 &  0.00110070 \\
         & $^6I$ & 13/2 & 0.02577610 & 0.02356034 &  0.00221576 \\
         & $^4I$ & 11/2 & 0.03030130 & 0.02794633 &  0.00235497 \\
         & $^6I$ & 15/2 & 0.03808675 & 0.03464574 &  0.00344101 \\
$4f^45d$ & $^6L$ & 11/2 & 0.04051482 & 0.04043802 &  0.00007680 \\
$4f^46s$ & $^4I$ & 13/2 & 0.04490881 & 0.04112088 &  0.00378793 \\
         & $^6I$ & 17/2 & 0.05111470 & 0.04634376 &  0.00477094 \\
$4f^45d$ & $^6L$ & 13/2 & 0.05192731 & 0.05000719 &  0.00192012 \\
         & ... & ... & ... & ... & ...  \\
$4f^35d^2$ & $^6M^o$ & 13/2 &0.15880928 & 0.07299076 & 0.08581852\\

\hline\hline
\end{tabular}
\end{table*}

\begin{table*}
\centering
\caption{Comparison of calculated energy levels of Nd {\sc iii} from AS in Ry with those available in NIST. Those levels denoted $^b$ were only available from \citet{Ding_Ryabtsev_Kononov_Ryabchikova_Clear_Concepcion_Pickering_2024}$^b$ (converted from cm$^{-1}$). All other levels were available in both sources and had good agreement between them.}
\renewcommand{\arraystretch}{1.2}
\label{Ndiii_levels}
\begin{tabular}{lcccccc}
\hline\hline
Configuration& Term& J&	$E_{calc}$ &	$E_{exp}$ &	$\Delta $ E  \\
\hline
$4f^4$ & $^5I$ & 4 & 0.000 & 0.000 &  0 \\
       & $^5I$ & 5 & 0.01044601 & 0.010368 & 0.00007801 \\
       & $^5I$ & 6 & 0.02205124 & 0.021757 & 0.00029424  \\
       & $^5I$ & 7 & 0.03448645 & 0.033853 & 0.00063346  \\
       & $^5I$ & 8 & 0.04748733 & 0.046414 & 0.00107333  \\
       & $^5F$ & 2 & 0.11782631 & 0.098179$^b$ & 0.01964731\\
       & $^5F$ & 3 & 0.12384761 & 0.104108$^b$ & 0.01973961\\
$4f^35d$ & $^5K^o$ & 5 & 0.14603556 & 0.139079  & 0.0069565  \\
         & $^5K^o$ & 6 & 0.16508120 & 0.154351 & 0.0107302  \\
         & $^5K^o$ & 7 & 0.18741103 & 0.170009 & 0.0174020  \\

\hline\hline
\end{tabular}
\end{table*}

\begin{table*}
\centering
\caption{Comparison of calculated energy levels of Nd {\sc iv} from AS in Ry with those available in \citet{wyart_analysis_2007} (converted from cm$^{-1}$) }
\renewcommand{\arraystretch}{1.2}
\label{Ndiv_levels}
\begin{tabular}{lcccccc}
\hline\hline
Configuration& Term& J&	$E_{calc}$ &	$E_{exp}$ &	$\Delta $ E  \\
\hline
$4f^3$ & $^4I^o$ & 9/2 & 0.000 & 0.000 &  0 \\
       & $^4I^o$ & 11/2 & 0.01371385 & 0.017287 & 0.0035731 \\
       & $^4I^o$ & 13/2 & 0.02852730 & 0.035607 & 0.0070797\\
       & $^4I^o$ & 15/2 & 0.04410995 & 0.054571 & 0.0104610\\
       & $^2H^o$ & 9/2 & 0.10714620 & 0.116644 &  0.0094978 \\
       & $^4F^o$ & 3/2 & 0.10927208 & 0.106604 &  0.0026680 \\
       & $^4F^o$ & 5/2 & 0.11675007 & 0.116167 &  0.0005831\\
       & $^4S^o$ & 3/2 & 0.12297689 & 0.125686 &  0.0027091 \\
       & $^4F^o$ & 7/2 & 0.12352500 & 0.125024 &  0.0014990 \\
       & $^2H^o$ & 11/2 & 0.13025935 & 0.147274 & 0.0170146  \\
       & ... & ... & ... & ... & ...  \\
$4f^25d$ & $^4K$ & 11/2 &  0.67071795 & 0.653784 & 0.0169339 \\
        & $^2H$ & 9/2  & 0.68954230 & 0.645333 & 0.044209\\

\hline\hline
\end{tabular}
\end{table*}

\begin{table*}
\centering
\caption{Comparison of calculated energy levels of Nd {\sc v} from AS in Ry with those available in \citet{meftah_spectrum_2008} (converted from cm$^{-1}$)}
\renewcommand{\arraystretch}{1.2}
\label{Ndv_levels}
\begin{tabular}{lcccccc}
\hline\hline
Configuration& Term& J&	$E_{calc}$ &	$E_{exp}$ &	$\Delta $ E  \\
\hline
$4f^2$ & $^3H$ & 4 & 0.000 & 0.000 &  0 \\
       & $^3H$ & 5 & 0.02707074 & 0.025828 & 0.00124274 \\
       & $^3H$ & 6 & 0.05496308 & 0.052337 & 0.00262608  \\
       & $^3F$ & 2 & 0.06503546 & 0.053707 & 0.01132846  \\
       & $^1G$ & 4 & 0.08177541 & 0.111813 & 0.03003759  \\
       & $^3F$ & 3 & 0.08267339 & 0.070942 & 0.01173139  \\
       & $^3F$ & 4 & 0.11779617 & 0.075740 & 0.04205617  \\
       & $^1D$ & 2 & 0.22524381 & 0.187278 & 0.03796581  \\
       & $^1I$ & 6 & 0.26989963 & 0.237732 & 0.03216763  \\
       & $^3P$ & 0 & 0.27872275 & 0.228278 & 0.05044475  \\
       & ... & ... & ... & ... & ... \\
$4f5d$ & $^1G^o$ & 4 & 1.16156315 & 1.162458 & 0.00089485 \\
       & $^3F^o$ & 2 & 1.16782919 & 1.162401 & 0.00542819 \\

\hline\hline
\end{tabular}
\end{table*}
The computation time was factored into the selection of configurations. Utilising the configuration averaged approach, a large set of candidate configurations was reduced after inspection of their contribution to the overall structure.
The structure for the Nd ions was optimized for the low-lying energy levels as an additional aid to reducing computation time and size.
The first $\sim$ten energy levels are given in tables \ref{Ndi_levels}-\ref{Ndv_levels}.The energy position of the first level with opposite parity to the ground is also given. These energies are those calculated in {\sc autostructure} as part of the structure calculations. The energies are shifted in the post-processing stage of recombination to the available experimental data and are shifted to the experimental data during the calculation of direct excitation. 
Given the temperature range of interest during the non-LTE phase of the kilonova, the detailed improvements of the structure can be limited to suit the relevant conditions.
Using the orbital,$\lambda$, scaling parameters, the structure was particularly sensitive to the $4f$-shell $\lambda$ parameter where small deviations caused energy level positions to move by up to one Rydberg ( 1 Ry $\approx 109,737.316$ cm$^{-1}$).
Nd {\sc iii} exhibited the strongest sensitivity, where small changes in the scaling parameter (at the level of the third decimal place) lead to substantial rearrangements of the level structure.
Through the optimization method, the overall atomic structure was achieved. The low-lying energy levels were further optimized with greater precision and accuracy, while less effort was devoted to the higher lying energy positions.
The precision of the high-lying levels was found to have little impact on the final rate and was therefore neglected.

The energy level positions were compared to available experimental data or other theoretical data when experimental validation was not available.
For all ion stages, NIST ASD was used as a source of data where data was available.
For Nd {\sc i} to {\sc iii} there were a reasonable number of energy levels \citep{Martin_Zalubas_Hagan_1978} to use as a benchmark.
Data from \cite{Ding_Ryabtsev_Kononov_Ryabchikova_Clear_Concepcion_Pickering_2024,ding_spectrum_2024} was also used to investigate energy level positions of Nd {\sc iii}.
\cite{wyart_analysis_2007} was used to determine the structure of the Nd {\sc iv} ion and \cite{meftah_spectrum_2008} was used to verify level positions of the Nd {\sc v} structure.

Calculated transition probabilities are compared in table \ref{A_vals} with the limited experimental values available in the literature \citep{Reader_Corliss_Wiese_Martin_1980,wyart_analysis_2007,meftah_spectrum_2008,Ding_Ryabtsev_Kononov_Ryabchikova_Clear_Concepcion_Pickering_2024}.
The comparison includes those transitions for which collision strengths are presented in section \ref{sec:ExcResults} and are denoted by a star ($\star$) within the table.

\begin{table}
\centering
\caption{Comparison of gA calculated by {\sc autostructure} with literature values. Those denoted $^a$ come from NIST\citep{Reader_Corliss_Wiese_Martin_1980}, $^b$ from \citet{Ding_Ryabtsev_Kononov_Ryabchikova_Clear_Concepcion_Pickering_2024}, $^c$ from \citet{wyart_analysis_2007} and $^d$ from \citet{meftah_spectrum_2008}. Transitions denoted with $\star$ have collision strengths presented in section \ref{sec:ExcResults}}
\renewcommand{\arraystretch}{1.2}
\label{A_vals}
\begin{tabular}{lccccc}
\hline\hline
Ion & Transition & gA ($s^{-1}$) & gA ($s^{-1}$) \\
 &  &  (AS) &  (Literature) \\
\hline
Nd I & $4f^46s^2 (^5I_5) -4f^46s6p (^5K_6^o)$ & 3.08E+09 & 7.70E+08 $^a$ \\
     & $4f^46s^2 (^5I_8) -4f^46s6p (^5K_9^o)$ & 5.57E+09 & 1.70E+09 $^a$ \\
     & $4f^46s^2 (^5I_4) -4f^46s6p (^5K_5^o)$ & 2.45E+09 & 9.90E+08 $^a$ \\
$\star$ & $4f^46s^2 (^5I_4) -4f^46s6p (^5H_3^o)$ & 9.45E+08 & 5.90E+08 $^a$  \\
\hline
Nd II & $4f^46s (^6I_{7/2}) -4f^46p (^4H_{7/2}^o)$ & 3.70E+07 & 4.50E+07 $^a$ \\
    &$4f^46s (^6I_{11/2}) -4f^46p (^6H_{11/2}^o)$ & 8.03E+08 & 1.58E+08 $^a$  \\
    &$4f^46s (^6I_{7/2}) -4f^46p (^6H_{7/2}^o)$ & 2.53E+07 &7.10E+07 $^a$ \\
    & $4f^45d (^6I_{11/2}) - 4f^46p (^6K_{13/2}^o)$ & 3.89E+07 & 2.07E+07 $^a$ \\
\hline
Nd III & $4f^3 5d (^5G_4^o) -4f^36p (^5G_5)$ & 1.67E+06 & 3.30E+06 $^b$   \\
 $\star$   & $4f^4(^5I_4) -4f^35d (^5K_5^o)$ & 9.80E+06 & 5.80E+06 $^b$   \\
    & $4f^4(^5F_4) -4f^35d (^5G_5^o)$ & 5.01E+06 & 3.80E+06 $^b$  \\
    & $4f^4(^5F_2) -4f^35d (^3G_3^o)$ & 1.90E+06 & 1.80E+06 $^b$ \\
    & $4f^35d(^5L_6^o) -4f^36p(^3I_5)$ & 6.09E+08 & 6.10E+08 $^b$ \\

\hline
Nd IV & $4f^3 (^2G_{7/2}^o) -4f^25d (^2G_{7/2})$ & 1.46E+08 & 1.86E+08 $^c$ \\
    & $4f^3 (^4F_{5/2}^o) -4f^25d (^4G_{7/2})$ & 9.20E+08 & 1.60E+08 $^c$ \\
    & $4f^3 (^4G_{9/2}^o) -4f^25d (^4G_{9/2})$ & 7.16E+07 & 1.09E+08 $^c$\\
    & $4f^3 (^4G_{9/2}^o) -4f^25d (^4H_{11/2})$ & 4.88E+08 & 1.33E+08 $^c$\\
    & $4f^3 (^4G_{7/2}^o) -4f^25d (^4G_{5/2})$ & 6.66E+06 & 8.00E+06 $^c$ \\
 $\star$   & $4f^3 (^4I_{9/2}^o) -4f^25d (^2H_{9/2})$ & 5.93E+08 & 5.37E+08 $^c$\\
\hline
Nd V & $4f^2 (^3H_4) -4f5d (^3D_3^o)$ & 1.37E+07 & 3.20E+07 $^d$ \\
    & $4f^2 (^3H_4) -4f5d (^1G_4^o)$ & 1.56E+09 & 4.13E+08 $^d$ \\
    & $4f^2 (^3F_4) -4f5d (^1G_4^o)$ & 3.98E+08 & 3.22E+08 $^d$  \\
    & $4f^2 (^3F_2) -4f5d (^3D_1^o)$ & 3.27E+09 & 6.97E+08 $^d$ \\
    & $4f^2 (^3P_2) -4f5d (^1F_3^o)$ & 3.09E+08 & 2.59E+08 $^d$ \\
  $\star$  & $4f^2 (^3H_4) -4f5d (^1G_4^o)$ & 1.11E+09 & 1.72E+09 $^d$ \\
\hline\hline
\end{tabular}
\end{table}

\begin{figure*}
    \centering
    \includegraphics[width=\linewidth]{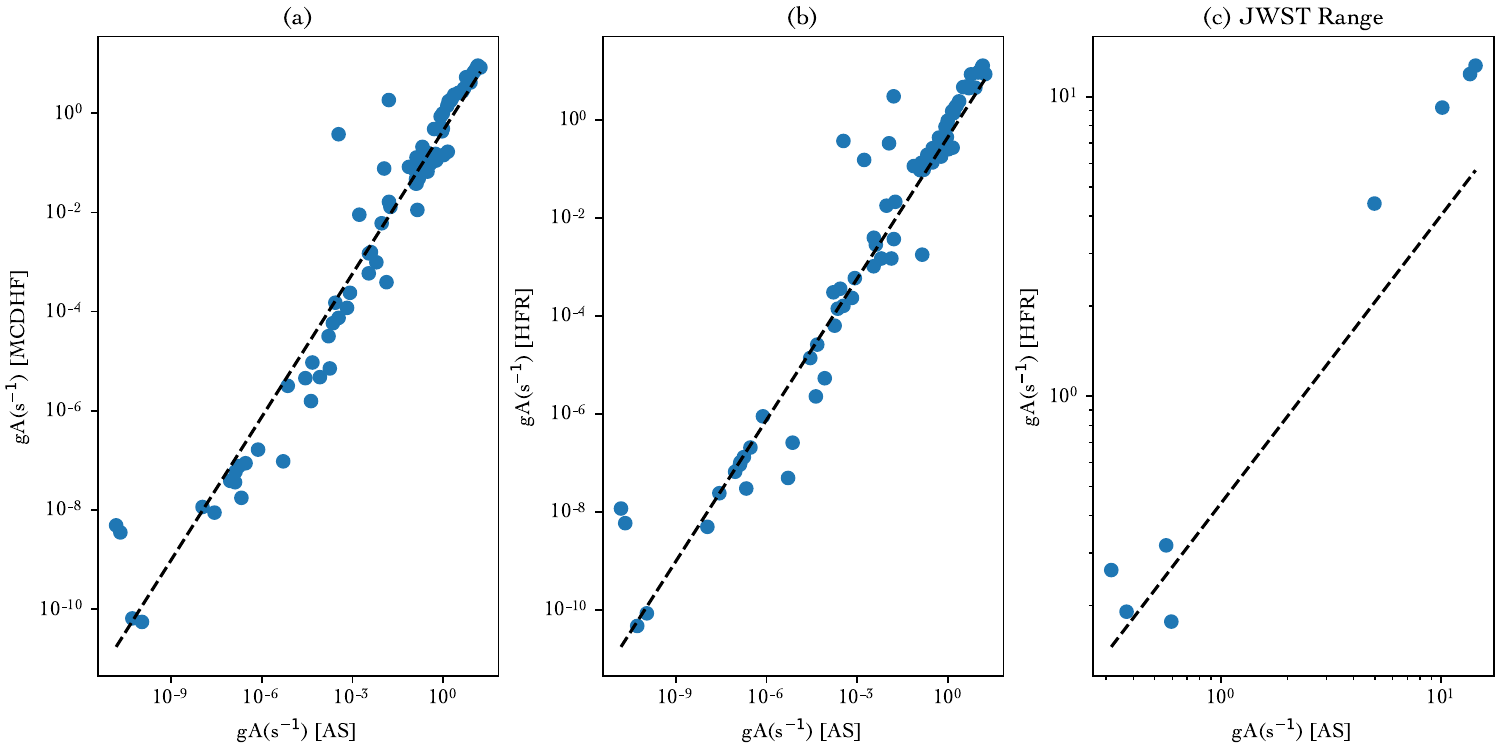}
    \caption{Comparison of weighted transition probabilities (gA) for forbidden transitions in Nd {\sc{iii}}. Each point represents a single transition with the {\sc autostructure} value on the horizontal axis and the corresponding \citet{maison_calculation_2024} value on the vertical axis (a) Comparison of gA-values from {\sc autostructure} against MCDHF method. (b) Comparison of gA-values from {\sc autostructure} against HFR method. (c) Comparison of 21 strongest lines discussed in \citet{maison_calculation_2024} that are within the observational range of JWST. The dashed line denotes perfect agreement.}
    \label{fig:III_A-vals}
\end{figure*}

Forbidden transition probabilities of Nd {\sc{iii}} are compared with values reported in the work by \cite{maison_calculation_2024} for a selection of transitions and table \ref{tab:NdIIIforbid} provides a comparison for the most intense forbidden transitions. 
While a wide range of transitions is examined to assess overall agreement, a smaller subset of lines of particular relevance for observational diagnostics is discussed in more detail.
The 78 transitions shown in Figure \ref{fig:III_A-vals} were taken from table 2 of \cite{maison_calculation_2024} which were derived from the 16 experimental energy levels of the $4f^4$ configuration measured by \cite{Ding_Ryabtsev_Kononov_Ryabchikova_Clear_Concepcion_Pickering_2024,ding_spectrum_2024}.
These forbidden transitions are of particular interest in low-density environments, where metastable levels can persist long enough to produce observable emission features relevant to late-time spectra.
Transition probabilities that lie within the wavelength range of the \textit{JWST Near Infrared Spectrograph} (0.6–5.3 $\mu$m) were compared with those 21 strongest transitions stated in \cite{maison_calculation_2024}.
Comparisons from {\sc autostructure} were made with the 78 gA-values obtained with both HFR and MCDHF methods presented in the aforementioned work and are shown in Figure \ref{fig:III_A-vals}.
Each point in Figure \ref{fig:III_A-vals} represents a single forbidden transition, with the {\sc autostructure} value plotted on the horizontal axis and the corresponding value from \cite{maison_calculation_2024} on the vertical axis.
Overall, reasonable agreement is observed between the calculations. Approximately 71\% of the MCDHF transition probabilities and 82\% of the HFR transition probabilities agree with the present calculation within a factor of five.
In particular, the strongest forbidden transitions that lie within the wavelength range of interest for JWST observations are reproduced well.
For example the $^5I_5-^3K_6$ (7736\AA), $^5I_6 - ^3K_7$ (7833\AA) and $^5I_7 - ^3K_7$ (8742\AA) transitions, provided in table \ref{tab:NdIIIforbid}, differ from the HFR calculations by 10-15\%.
The remaining discrepancies are primarily associated with relatively weak transitions (\textit{gA}$\leq 10^{-4}s^{-1}$) which are known to be particularly sensitive to the details of the atomic structure.

\begin{table}
	\centering
	\caption{Most intense forbidden transitions of Nd III involving the $4f^4$ levels lying below the opposite parity and in the wavelength range that may be detected by JWST compared with \citet{maison_calculation_2024}.}
	\label{tab:NdIIIforbid}
	\begin{tabular}{lccr} 
		\hline
		$\lambda$ (\AA) & Transition & gA (HFR) & gA (AS)\\
         &  & \citet{maison_calculation_2024} & \\
		\hline
    6514.827  & $^5I_4 - ^5G_3$ & 1.91E-01 &3.72E-01\\
       7036.403 & $^5I_5 - ^5G_3$ & 1.77E-01 &5.94E-01\\
        7736.065& $^5I_5 - ^3K_6$ & 9.20E+00 &1.01E+01\\
        7833.146& $^5I_6 - ^3K_7$ & 1.27E+01 &1.43E+01\\
        8209.286& $^5I_4 - ^5F_4$ &2.63E-01 &3.17E-01\\
        8564.039& $^5I_6 - ^3K_6$ & 4.40E+00 &4.97E+00\\
        8741.825& $^5I_7 - ^3K_7$ & 1.19E+01 &1.35E+01\\
        21854.871& $^5F_2 - ^5G_3$ & 3.18E-01 &5.63E-01\\
		\hline
	\end{tabular}
\end{table}

\subsection{{\sc grasp}$^0$ : Atomic structure}
The general-purpose relativistic atomic structure package ({\sc grasp}$^0$) \citep{Grant80,Dyall1989} was used to obtain a structure for the Nd {\sc ii} ion which was subsequently used in an electron-impact excitation calculation using the Dirac Atomic R-matrix Codes ({\sc DARC}) \citep{norrington_1987,ParallelRmatrix}. {\sc grasp}$^0$ uses a Multi-Configurational Dirac-Fock (MCDF) method to generate electron orbitals to calculate a structure. Within a given configuration set, radial orbitals are determined by a self-consistent multi-configuration-Dirac-Hartree-Fock method. The configurations used are shown in table~\ref{tab:configs}. The additional configurations in the {\sc grasp}$^0$ calculation that were not included in the {\sc autostructure} were added to aid in converging the calculation, those configurations are not expected to be relevant at the temperature ranges used for modelling in this paper.

The {\sc DARC} is a suite of codes developed to model varied atomic processes, based on the R-matrix theory \citep{Burke_2011}. In this work the {\sc grasp}$^0$ atomic structure calculation for Nd {\sc ii} was used as the basis for an electron-impact excitation calculation, to compare to the distorted wave calculation done using {\sc autostructure}. This structure calculation used the configuration set shown in table~\ref{tab:configs}.

 \section{Recombination}
 \subsection{Theory}
In the isolated-resonance approximation adopted in this work, recombination is determined by the underlying atomic structure through the calculation of autoionization and radiative rates. Consequently, accurate atomic structure calculations are essential for obtaining reliable rate coefficients.
Alternative formulations, such as the unified recombination method \citep{Nahar_Pradhan_1992,Nahar_Pradhan_1994}, treat radiative and dielectronic recombination self-consistently within a single framework.

The dominant recombination process is dielectronic recombination (DR) which is generally dominant over radiative recombination (RR) for those ions with fine-structure.
DR is a two-step process involving the capture of a continuum electron, resulting in an ion core excitation with subsequent radiative stabilisation of the core, 
\begin{equation}
    A^{q+}+e^- \leftrightarrow A^{**(q-1)+} \rightarrow A^{(q-1)+} + \gamma.
\label{eq:DRProcess}
\end{equation}
The captured electron occupies a highly excited level of the ion, and the ion stabilises in one of two ways.
The system may either autoionize back to the continuum, or radiative decay to a true bound state.
The DR rate coefficient is computed by,
\begin{equation}\label{eq:DRRate}
\begin{split}
    \alpha_{DR}(i+e^- &\rightarrow f) = {\left({\frac{4\pi{\alpha_0}^2 I_H}{kT}}\right)}^{\frac{3}{2}} \sum_{j}\frac{g_j}{2g_i}e^{{-E}/{kT}} \\
    & \times \frac{\sum_l A_a(j\rightarrow i,kl) A_r(j \rightarrow f)}{\sum_h A_r(j\rightarrow h) + \sum_{m,l} A_a(j\rightarrow m,kl)},
\end{split}
\end{equation}
with $i$ and $f$ the initial and final states, $j$ the doubly excited intermediate state, $E$ the resonance energy, $g_i$ the statistical weight of the \textit{N}-electron target ion, $g_j$ the statistical weight of the (\textit{N}+1)-electron intermediate state and $T$ is the temperature (K). 
The dielectronic capture rate coefficient is written in terms of the autoionization rate, using the principle of detailed balance assuming a Maxwellian distribution, multiplied by the fraction that stabilise.
Resonances are treated within the isolated-resonance approximation and assumed to have Dirac delta-function profiles in energy, valid when their natural widths are small compared to the thermal energy spread of the electrons.

RR is notably the initial and final stages of DR.
\begin{equation}
    A^{q+}+e^- \rightarrow A^{(q-1)+} + \gamma.
\end{equation}
The electron is captured by a charged ion and simultaneously emits a photon such that the final state is bound.
The radiative rate coefficient can be computed via, 
\begin{equation}
\label{RRRate}
\begin{split}
     \alpha_{RR}(i\rightarrow f&;T_e)=\left(\frac{2a_0^2}{\tau_0\sqrt{\pi}}\right)\left(\frac{I_H}{kT_e}\right)^{3/2}\\
   &\int^\infty_0 {\frac{\varepsilon_i}{I_H}e^{-\varepsilon_i/{kT}}\sigma^{RR} \frac{d\varepsilon_i}{I_H}}  
\end{split}
\end{equation}
where, ${{2a_0}/{\tau_0\pi^{1/2}}=2.46854\times10^8}$cm/s, $T$ is temperature, $\varepsilon_i$ is the energy of the incident electron, $k$ is the Boltzmann constant and 
\begin{equation}
\sigma^{RR}(i+\varepsilon_i\rightarrow f+\gamma)=\frac{\pi a_0^2\alpha^3E^3_{\gamma}}{3g_i\varepsilon_i I_H^2}S(i\rightarrow f)
\end{equation}
is the RR cross-section with photon energy $E_{\gamma}$, $S(i\rightarrow f)$ the electric dipole line strength for the bound-free transition, evaluated in the length gauge and $\pi a_0^2\alpha^3/{3}=1.1395\times10^{-23}$cm$^2$.
The RR cross-sections are obtained via detailed balance from computing photoionization cross-sections for bound levels.
Consequently, photoionization cross sections are computed simultaneously with the recombination calculation.
The total recombination rate coefficient, as used in many modelling codes, is obtained by summing the final temperature dependent RR and DR rate coefficients.

\subsection{Recombination Results} \label{sec:RecResults}
For the DR calculations, each configuration in Table~\ref{tab:configs} is augmented by a Rydberg electron characterized by quantum numbers $n\ell$, with $n$ principal quantum number and $\ell$ the angular momentum quantum number.
DR from Rydberg $n\ell$ electrons are calculated explicitly up to $n=35$.
For higher $n$, a set of values between $n=35$ and $n=999$ is computed explicitly and the remaining intermediate $n$ values are obtained through interpolation.
For these ion stages, $\ell$ values are chosen to ensure convergence of the rate coefficient, typically with a maximum of $\ell=10$, except for Nd {\sc ii}.
Due to the computational demand of Nd {\sc ii}, Rydberg $n\ell$ were explicitly computed up to $n=14$ and $\ell=4$.

The RR rate coefficients were computed in a similar fashion. Although, fewer explicit $\ell$ values were required to achieve convergence.
The total RR rate coefficients were calculated from the direct photo-ionization cross sections using detailed balance.
The post-processors {\sc adasdr} and {\sc adasrr} were used to process the radiative rates, autoionization rates and photoionization cross-sections. 
During the post-processing, high-$\ell$ states are treated using the hydrogenic approximation in {\sc adasrr}. For DR, transition probabilites between high-$n$ Rydberg states are evaluated hydrogenically within {\sc adasdr} using hydrogenic dipole radial matrix elements calculated by the Burgess recurrence algorithm \citep{Burgess_1965,burgess_general_1965}.

\begin{figure*}
    \centering
    \includegraphics[width=0.49\linewidth]{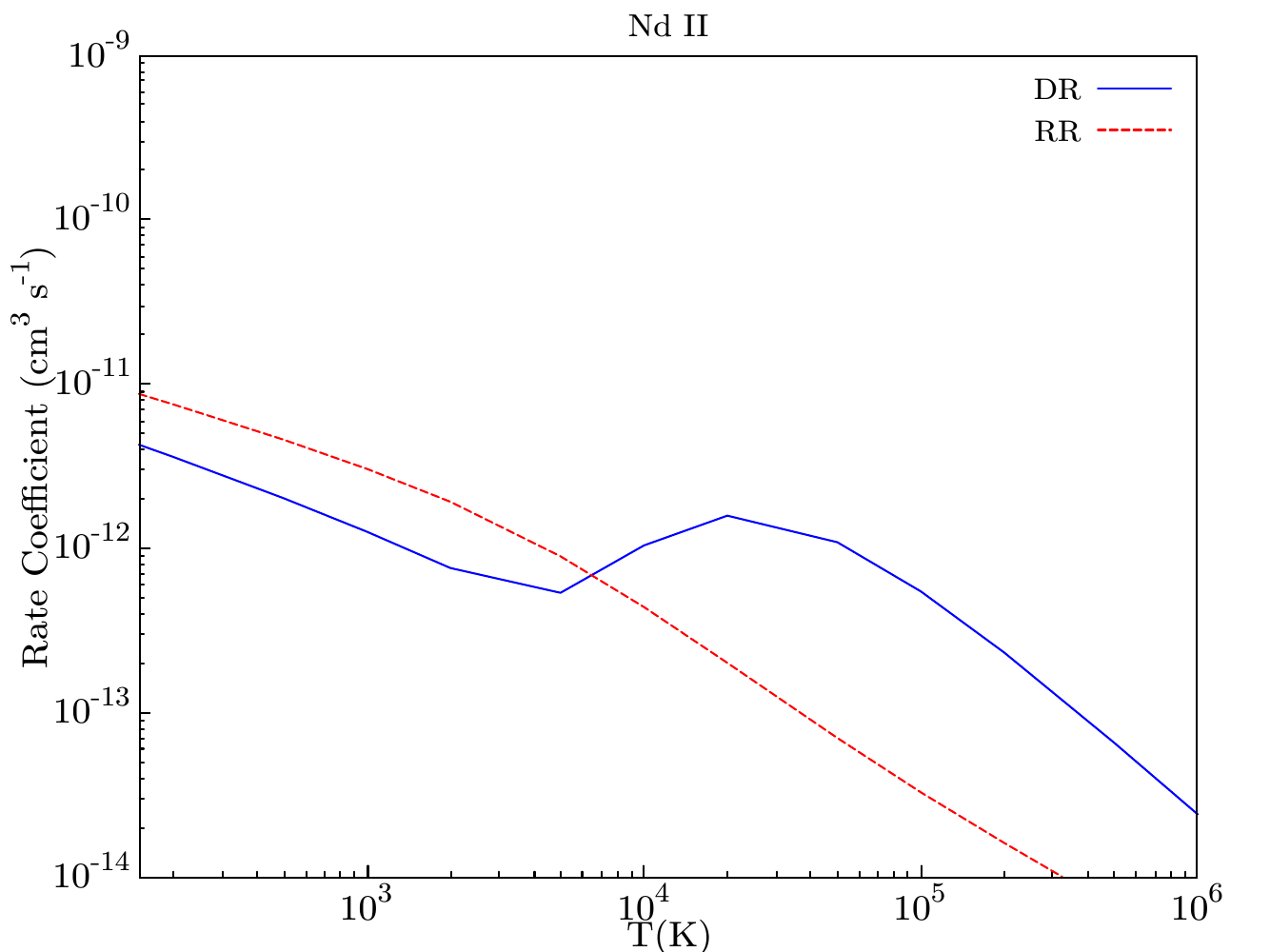} 
    \includegraphics[width=0.49\linewidth]{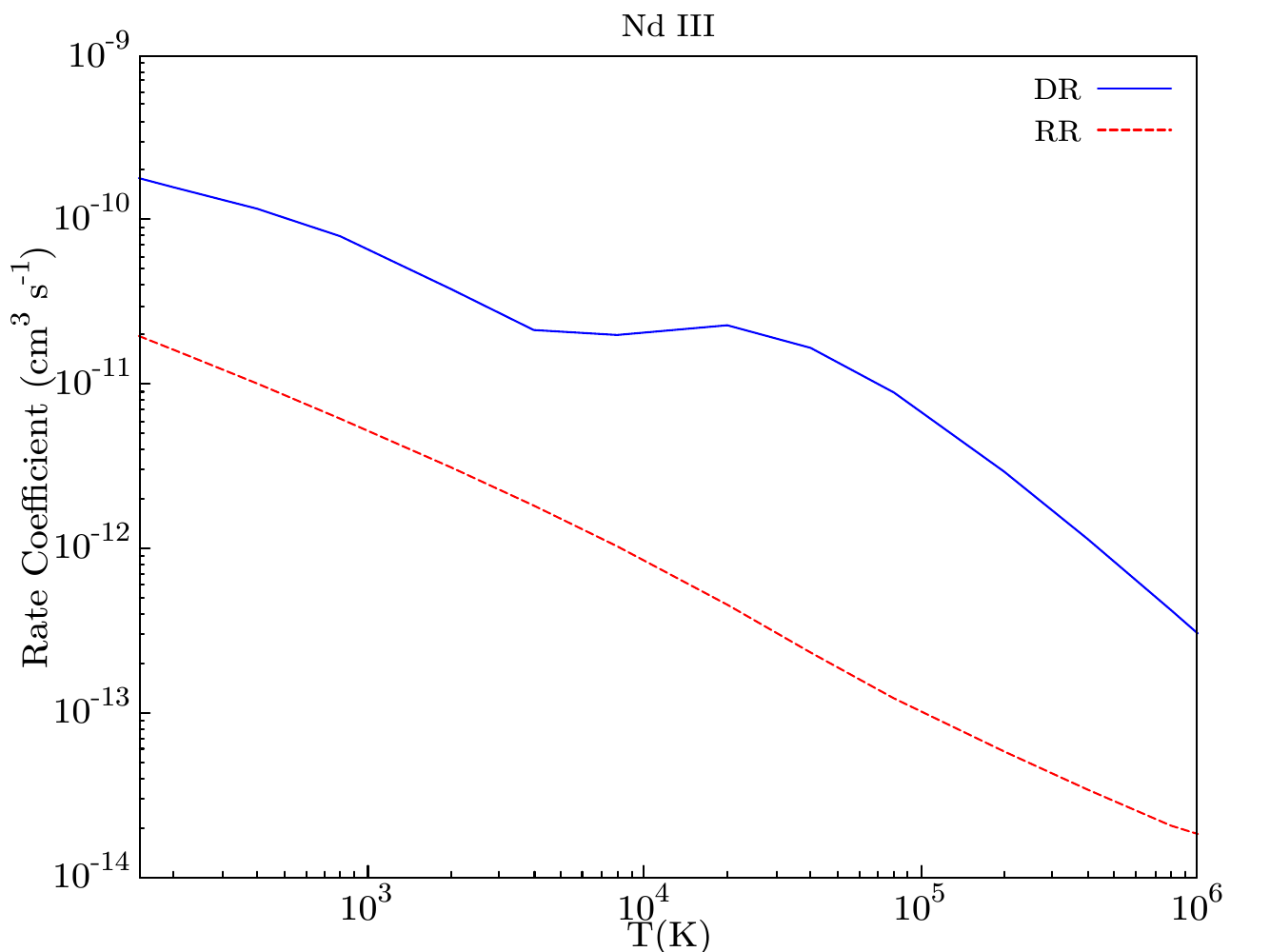} \\
    \includegraphics[width=0.49\linewidth]{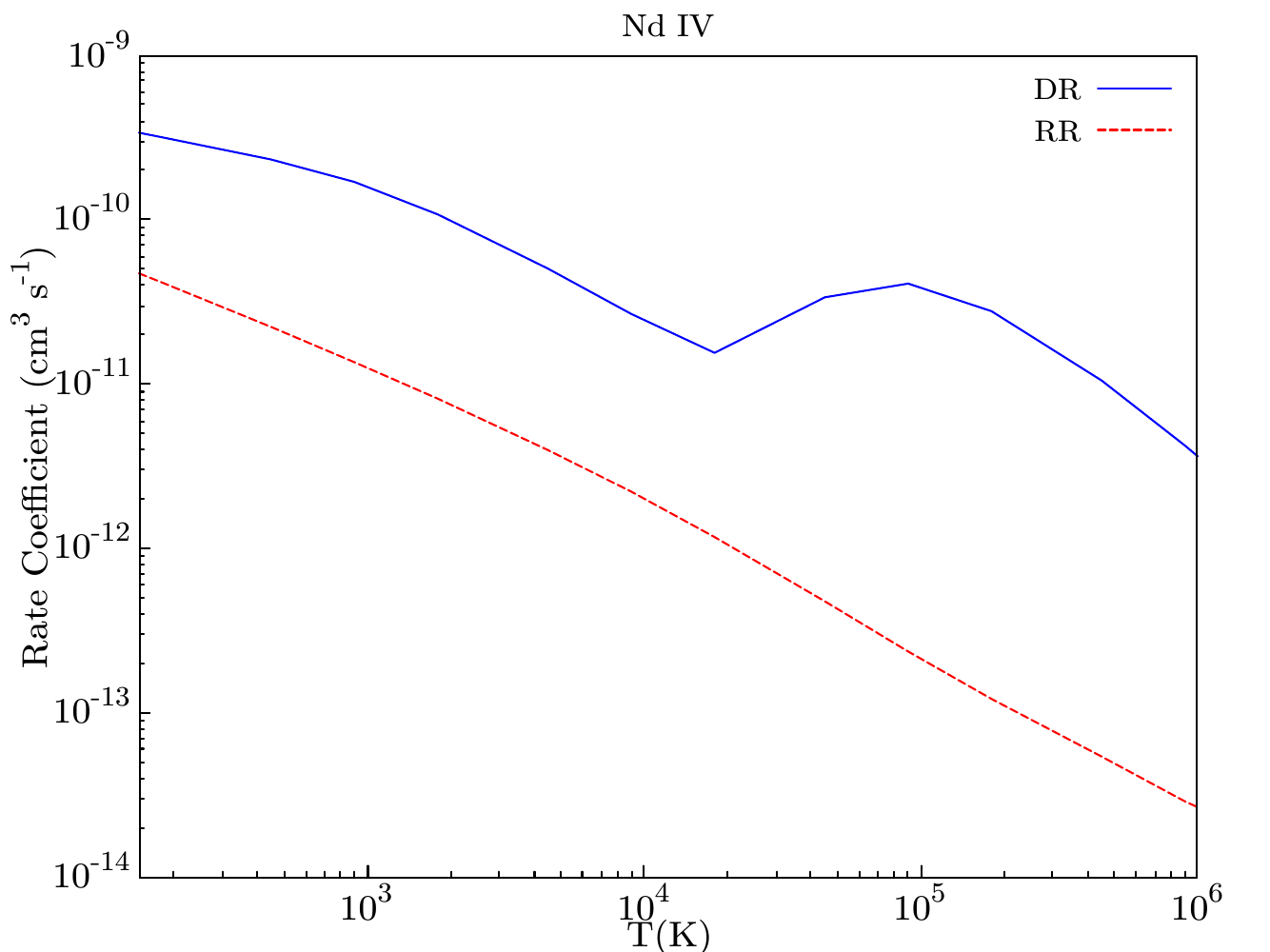} 
    \includegraphics[width=0.49\linewidth]{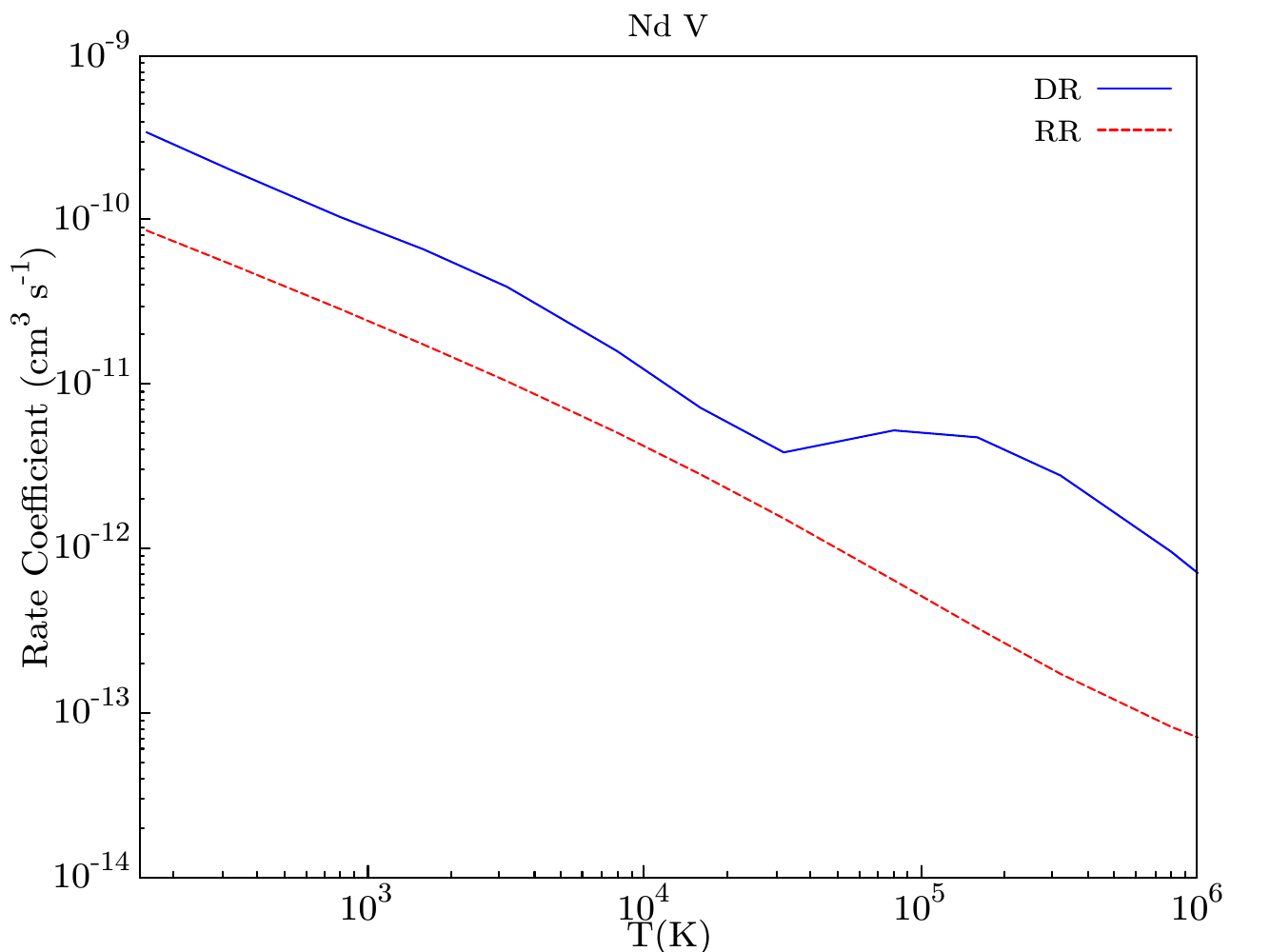} 
    \caption{Radiative and dielectronic recombination rate coefficients for Nd $II\rightarrow I$ [top left], $III\rightarrow II$ [top right], $IV \rightarrow III$ [bottom left], $V \rightarrow IV$ [bottom right]}
    \label{fig:Recombination}
\end{figure*}

The level-resolved recombination rate coefficients for Nd ions {\sc ii - v} are shown in Figure \ref{fig:Recombination}.
It is evident for ions {\sc iii}-{\sc v} that the DR is the dominant recombination process within the nebular phase temperatures.
However, it is clear that the RR of {\sc ii}$\rightarrow$ {\sc i} is greater than the dielectronic at low temperatures.
In general, these recombination rate coefficients are higher than the analytical rate coefficient of $10^{-11}\rm{cm}^3\rm{s}^{-1}$ which is currently being used within some models (eg. \cite{pognan_validity_2022,pognan_nlte_2022} for ions where no data is currently available.
This approximation can strongly underestimate the contribution from recombination and skew model spectra.

The recombination rates of {\sc v} $\rightarrow$ {\sc iv}, {\sc iv} $\rightarrow$ {\sc iii} and {\sc iii} $\rightarrow$ {\sc ii} have also been computed by \cite{hotokezaka_nebular_2021} using {\sc hullac} and so comparisons can be drawn.
The DR rate coefficients computed by {\sc autostructure} feature a high temperature peak which one normally expects.
While efforts were made to ensure the overall high-energy structure is reasonable, greater emphasis was placed on the low-energy regime (as discussed in section \ref{sec:structure}) due to its relevance for neutron star merger applications; consequently, the high-temperature DR presented here should be interpreted with some caution.

\subsubsection{Nd II $\rightarrow$ I}
The ground level of Nd {\sc ii} is $^6I_{7/2}$ with configuration of $[Xe] 4f^4 6s$. 
NIST ASD has 840 levels within the database for Nd II.
In obtaining the structure, {\sc autostructure} computed 6888 levels from the configuration set given in table \ref{tab:configs}.

\begin{figure}
    \centering
    \includegraphics[width=\linewidth]{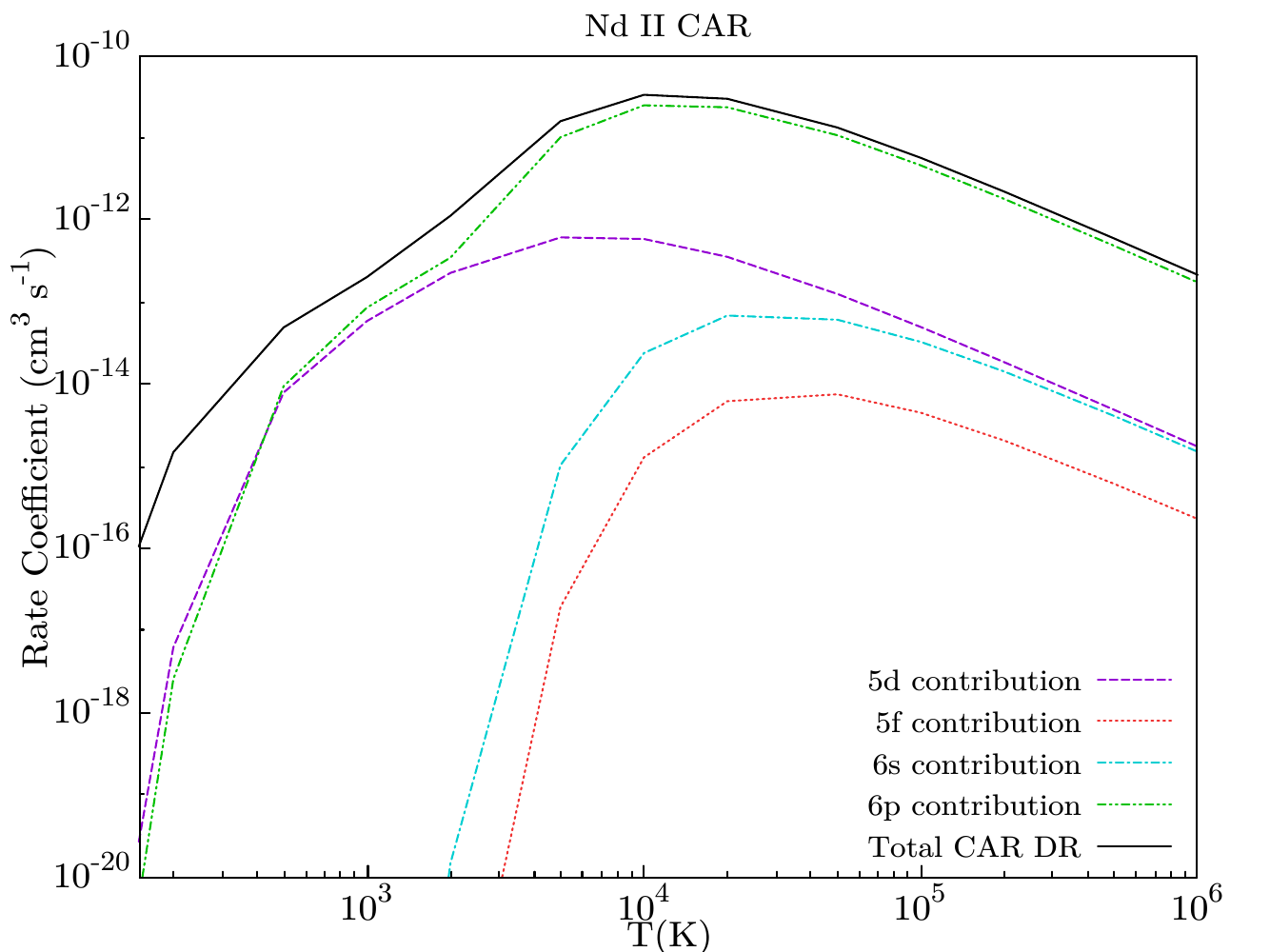}
    \caption{Nd II$\rightarrow$I DR using relativistic configuration average to identify the individual capture channels to the total DR and define a reduced model for the subsequent intermediate coupling calculation. }
    \label{fig:Recomb_II_CAR}
\end{figure}

\begin{figure}
    \centering
    \includegraphics[width=\linewidth]{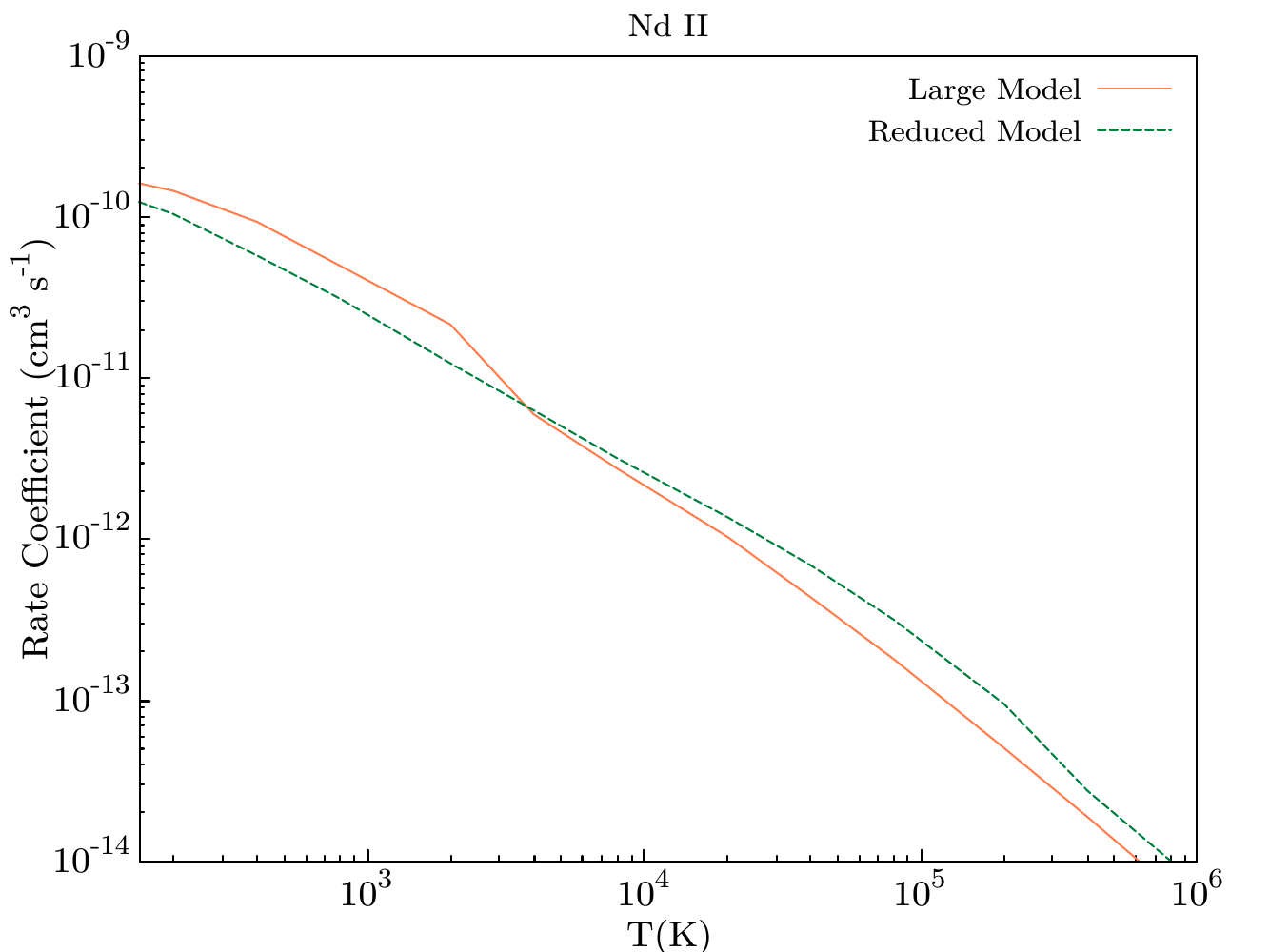}
    \caption{Comparison of core runs computed using the full $n\ell$ configuration set (orange) and the reduced configuration set (green) in intermediate coupling. The core calculation only includes the target core excitations, excluding the Rydberg $n\ell$ electron.}
    \label{fig:Core_comparison}
\end{figure}

Obtaining the recombination data for Nd {\sc ii $\rightarrow$ i} was computationally expensive.
The calculation was ultimately restricted to valence orbitals of $4f,5d,6s,6p$ to reduce the size of the computation.
However, an initial structure and DR calculation was completed using the configuration average coupling scheme with a complete $n\ell$ configuration set to investigate which configurations were contributing to the total DR rate coefficient.
The complete configuration set contained orbitals $ns,np,nd,nf,ng$ for $n= $ 5 and 6.
The relativistic configuration average DR is shown in Figure \ref{fig:Recomb_II_CAR}. The $5f$ and $6s$ contributed very little to the DR along with the $6d,6f,6g$ orbitals.
This calculation led to the identification of a reduced configuration set which was subsequently used in the recombination calculations.
Regardless, an intermediate coupling core run (ie. no Rydberg states included) was calculated with the complete $n\ell$ configuration set to check which resonances drove the recombination.
The core run from the complete $n\ell$ model was compared with the core run from the reduced configuration set to give reassurance that essential resonances contributing to the DR rate coefficient were not missed by using the reduced configuration set.
From Figure \ref{fig:Core_comparison}, there is little difference between both core runs which ensures no crucial resonances are being missed by using the smaller configuration set.
The configuration-averaged calculation shown in Figure \ref{fig:Recomb_II_CAR} is used solely as a computationally inexpensive screening tool to identify the dominant recombination channels and construct a reduced configuration set. Since configuration average neglects fine-structure mixing present in the subsequent intermediate coupling calculations, quantitative agreement with the intermediate coupling rate coefficients shown in Figure \ref{fig:Recombination} is not expected. 
The validity of the reduced configuration set is instead established by the comparison of full and reduced intermediate coupling calculations presented in Figure \ref{fig:Core_comparison}.

This provided confidence on continuing with the restricted configuration set that could be used to compute recombination rate coefficients within a reasonable computation time.
The final DR rate coefficient was computed with the reduced set of target configurations stated in table \ref{tab:configs}, shown in the upper left of Figure \ref{fig:Recombination}.
This DR rate coefficient was limited to Rydberg $l=0-4$ due to the computational demand of the ion stage.
Within the post-processing stage, some energy levels from NIST ASD were read in to further ensure correct resonance position. 
For Nd {\sc ii} $\rightarrow$ {\sc i}, levels up to 0.15 Ry had final adjustments through this shift.

At low temperatures, the RR rate coefficient becomes larger than the DR but with little difference between them.
It may be considered that the low temperature DR would become greater than the RR should more Rydberg $\ell$ be included but a significant change would generally not be expected.
RR dominates at low temperatures due to the limited number of low-lying autoionizing states in the parent ion, which suppresses DR until higher temperatures where additional core excitations become energetically accessible.

\subsubsection{Nd III $\rightarrow$ II}
The ground level of Nd {\sc iii} is $^5I_4$ with configuration of $[Xe] 4f^4$ as verified by NIST ASD where 30 energy levels are provided.
In total, {\sc autostructure} computed 817 energy levels within the structure.
Obtaining a good structure for Nd {\sc iii} came with challenges due to the sensitivity of the structure on the $4f$ orbital.
Minor adjustments within the third decimal point of the $\lambda$ scaling parameter, for the $4f$ orbital only, was the difference between obtaining the correct ground position and not.
As with each ion stage, an initial 'out-of-box' structure was computed using all default values ($\lambda = 1$) to investigate the overall level positions.
The out-of-box structure had an incorrect ground configuration, $4f^35d $ $^{5}L_6^{\rm{o}}$, and had extremely low energy positions for the excited levels, with the first excited level of $^{5}L_7^{\rm{o}}$ at 0.0026 Ry.
The optimized structure has adjusted $\lambda$ scaling parameters for all orbitals, not just $4f$, to ensure an overall good structure.

The DR rate coefficient shown in the top right of Figure \ref{fig:Recombination} was obtained using the optimized structure which had further fine-tuning up 0.1 Ry of the energies in the post-processing stage.
The energies used to verify the structure and during the final fine-tuning were those experimentally obtained from \cite{Ding_Ryabtsev_Kononov_Ryabchikova_Clear_Concepcion_Pickering_2024}.

\subsubsection{Nd IV $\rightarrow$ III}
The ground level of Nd {\sc iv} is $^4I^o_{9/2}$ with configuration of $[Xe] 4f^3$.
The number of energy levels provided by NIST ASD become limited at the Nd {\sc iv} stage where only 19 levels are given compared to the 241 computed in the structure run.
However, some experimental and theoretical data from \cite{wyart_analysis_2007,wyart_theoretical_2008} was utilised to qualify the structure.

Both RR and DR rate coefficients for Nd {\sc iv}$\rightarrow${\sc iii} are shown in the bottom left plot of Figure \ref{fig:Recombination}.
For the final fine-tuning of the energy levels in the post-processor, the values from NIST ASD are interpolated energy levels but alongside the data from \cite{wyart_analysis_2007,wyart_theoretical_2008}, the final adjustments to resonance positions were made up to 0.13 Ry.

In rare, less dense, plasma, the contribution to DR from higher $n$ must be accounted for. The rate coefficients shown in Figure \ref{fig:Recombination}, emphasises the importance that the high-n contribution has to the high temperature features.

\subsubsection{Nd V $\rightarrow$ IV}
The ground level of Nd {\sc v} is $^3H_4$ with configuration of $[Xe] 4f^2$.
Energy level position verification was conducted using experimental data from \cite{meftah_spectrum_2008} as the NIST ASD data only had one energy level, the ground.
The optimized structure had 69 energy levels in which they were optimized based on the few from \cite{meftah_spectrum_2008}.

Since only the ground level is stated in NIST ASD, experimental validation by \cite{meftah_spectrum_2008} was used for the final post-processor adjustments up to 0.1 Ry.
The value of the internal optimization workflow becomes particularly significant where little external benchmarking data is available, as it provides a systematic means of improving the consistency and reliability of the atomic structure in cases such as Nd {\sc v}.
The DR and RR rate coefficients are shown in the bottom right plot of Figure \ref{fig:Recombination}.

 \section{Excitation}
 
 \subsection{Theory}
Electron-impact excitation occurs when an incident electron has energy exceeding the excitation threshold (the energy difference between the initial and final states), and is characterised by the excitation cross section or, equivalently, the collision strength, $\Omega$.
Collision strengths are computed in a Breit–Pauli intermediate-coupling scheme (with the exception of the {\sc DARC} calculations, which are performed in $jjJ$-coupling) and are evaluated at the same set of final energies for all transitions.
The collision strength, $\Omega_{ij}$, varies with incident electron energy and is related to the cross-section by, 
\begin{equation}
    \Omega_{ij}=g_i(E_i/I_H)(\sigma_{i\rightarrow j}(\varepsilon_i)/\pi a_0^2)
\label{eq:collisionstrengths}
\end{equation}

where $\sigma_{i\rightarrow j}$ is the excitation cross-section, $\pi a_0^2$ is the atomic unit of cross-section,$\varepsilon_i$ incident electron energies, and $g_i$ are statistical weights.

The collision strengths are integrated over a Maxwellian distribution of electron energies at a set of electron temperatures to obtain the effective collision strengths $\Upsilon_{ij}$ given by,
\begin{equation}
    \Upsilon_{ij} =\int_0^\infty \Omega_{ij}(\varepsilon_j) \exp(-\varepsilon_j/kT_e) d(\varepsilon_j/kT_e)
\label{eq:Effcollisionstrength}
\end{equation}
which are related to the rates of excitation and de-excitation given by,
\begin{equation}
\begin{split}
q_{i\rightarrow j}(T_e) &= 2\sqrt{\pi}\alpha c a_0^2 
\frac{1}{g_i}
\left(\frac{I_H}{kT_e}\right)^{1/2}
\exp\left(-\frac{\Delta E_{ij}}{kT_e}\right)
\Upsilon_{ij},
\\
q_{j\rightarrow i}(T_e) &= 2\sqrt{\pi}\alpha c a_0^2 
\frac{1}{g_j}
\left(\frac{I_H}{kT_e}\right)^{1/2}
\Upsilon_{ij}.
\end{split}
\label{eq:excitationrate}
\end{equation}
where $2\sqrt{\pi}\alpha ca_o^2 = 2.1716\times 10^{-8} cm^3s^{-1}$. 

 \subsubsection*{Direct Excitation Theory}
{\sc autostructure} computes direct electron-impact excitation within the distorted-wave approximation, in both LS-coupling and level-resolved (intermediate coupling) formulations, the latter being employed in this work. 
In this approach, the scattering problem is treated using a distorted-wave representation of the incident and scattered electron, neglecting channel coupling but retaining the dominant direct excitation contributions. 
Collision strengths are computed for all transitions on a grid of threshold scaled units. These are then interpolated onto an absolute energy grid before being averaged over a Maxwellian distribution, to yield effective collision strengths for use in plasma modelling.
The electron-impact excitation rate coefficients computed here are useful in determining populations within a charge state.

\subsubsection*{Resonant Excitation Theory}
Resonant excitation (RE) can be computed alongside DR, as both processes rely on the same underlying atomic structure and share the initial step of dielectronic capture into auto-ionizing states.
Following this capture, RE proceeds via Auger decay back to the continuum, populating an excited state of the target ion, whereas DR proceeds via radiative decay to a bound state.
The atomic data generated can therefore be used to derive both RE and DR through appropriate post-processing procedures. Let $E_{ij} = E_j-E_i$ be  the energy difference between two states. Then, following \cite{sampson_fully_2009}, the resonant contribution to a collision strength is given by,
\begin{equation}
    \Omega^{\text{res}}_{if} = \frac{1}{2}h \sum_j \frac{g_j \sum_{l}A_a({j \to i},kl) \sum_{l'}A_a({j \to f},kl')}{\sum_{m,l} A_a({j \to m,kl})} \delta(\epsilon - E_{ij}), \label{eq:re_coll}
\end{equation}
where the quantities $A_a$ are Auger rates as discussed before. This equation encapsulates the dielectronic capture from state $i \to j$, followed by the autoionization  from state $j \to f$. Applying Equation \eqref{eq:Effcollisionstrength}, we have the contribution to the effective collision strength as,
\begin{equation}
    \Upsilon^{\text{res}}_{if} = \frac{1}{2}\frac{h}{kT} \sum_j \frac{g_j \sum_{l}A_a({j \to i},kl) \sum_{l'}A_a({j \to f},kl')}{\sum_{m,l} A_a({j \to m,kl})}  e^{ -{E_{jf}}/{kT}} \label{eq:re_eff}.
\end{equation}
The total effective collision strength is then the sum of this component with that of the direct-excitation discussed previously. This theory has been implemented in a post-processing code {\sc adasre}\footnote{Available at \emph{https://github.com/LeoMul/adasre}.}. 

\subsubsection*{R-Matrix method}
As will be discussed in section \ref{sec:ndIIexc}, the Dirac atomic R-matrix codes ({\sc DARC}) were used to compute electron-impact excitation for Nd {\sc ii}. Detailed theory can be found in \cite{Burke_2011} which has been applied in the parallel version \citep{ParallelRmatrix}.
The R-matrix method is a close-coupling approach for calculating electron-impact excitation cross sections in which configuration space is divided into an inner and an outer region. In the inner region, the scattering electron strongly interacts with the target ion and electron exchange and correlation effects are treated explicitly through a multi-configuration expansion of the total wavefunction. At the boundary of this region, the R-matrix provides the connection to the outer region, where the electron moves in the long-range potential of the target and the scattering problem is solved with appropriate asymptotic boundary conditions. This method naturally accounts for channel coupling and resonance structures, which are critical for accurate excitation rates, particularly for complex ions.

 \subsection{Excitation Results}\label{sec:ExcResults}

{\sc autostructure} has been used to compute direct excitation across the ion stages presented.
Computing excitation using the distorted wave approach implemented in {\sc autostructure} \citep{badnell_breitpauli_2011} enables a consistent structure to be used in all processes calculated.
However, for the case of Nd {\sc ii}, three excitation models have been calculated to discuss the different approaches to computing excitation and the differences that arises from such methods.
Resonances are calculated as part of the DR process and so can be used in addition to the distorted wave excitation to obtain the RE.
As discussed in section \ref{sec:ndIIexc}, including resonances within the excitation calculation can act as an alternative calculation to large R-Matrix computations that can be time consuming. This is particularly attractive when the required data will be calculated for the DR process in either case. The impact of the resonant contribution to a selection of dipole-allowed transitions are shown on Figures \ref{fig:Ndall_omega_res} (collision strengths) and \ref{fig:Ndall_ups_res} (effective collision strengths). It can be seen the contribution is primarily at low energy/temperatures. In the high energy/temperature regime, the resonances decay in magnitude - and the asymptotic behaviour of the dipole collision strength remains. In some of the cases shown, the low energy behaviour is quite pronounced. This behaviour is primarily due to very strong near-threshold resonances, particularly from the 4f$^k$ 5d$^1$ $nl$ set of Rydberg series. These resonances appear to attach very closely to the upper-level threshold, and combined with strong Auger rates resulting in a large effective collision strength at low electron temperatures. This additionally means that those collision strengths will be particularly sensitive to resonance positions in comparing calculations. Additionally, it is clear for a complex system such as Nd - that the standard high temperature behaviour of the collision strengths of dipole transitions can be potentially outweighed by resonant contributions at relatively high temperatures.

Although limited to the few ion stages of Nd here, further work on the validity of adding resonances to the distorted wave calculation will be explored with other species in upcoming work (Mulholland et al. {in prep.}).

\subsubsection{Nd I}
Neutral Nd has a reasonable number of experimentally verified energy levels in the NIST ASD database \citep{Martin_Zalubas_Hagan_1978}. There are noted to be 739 levels available whereas {\sc autostructure} obtains 3669 levels.
The configuration set was reduced for computation.
Direct excitation is computed for Nd {\sc i} using this structure with the first 25 energy levels shifted to match those up to 0.1 Ry from NIST ASD.
However, the distorted wave approach is less accurate for neutral atoms but is presented in table \ref{tab:NdI} to have a complete data set across the ion stages.
\begin{table*}
\centering
\caption{Effective collision strengths for the first five fine structure levels of Nd I. A full dataset will be available in the Zenodo repository and at Open-ADAS.}
\label{tab:NdI}
\renewcommand{\arraystretch}{1.2}
\begin{tabular}{lcccccccccc}
\hline\hline
& \multicolumn{10}{c}{Temperature (K)} \\
&2.00E+02& 5.00E+02& 1.00E+03&	2.00E+03 &	5.00E+03 &	1.00E+04 &	2.00E+04	& 5.00E+04 &	1.00E+05 &	2.00E+05 \\
\hline
Transition & \multicolumn{10}{c}{Effective collision strengths} \\
$^5I_4-^5I_5$ & 1.91E-03 & 4.64E-03 & 8.17E-03 & 1.30E-02& 2.36E-02 & 5.50E-02 & 3.21E-01 & 4.97E+00 & 1.55E+01 & 2.31E+01 \\
$^5I_4-^5I_6$ & 4.31E-05 & 1.07E-04 & 2.08E-04 & 3.98E-04& 9.33E-04 & 2.75E-03 & 8.36E-02 & 3.08E+00	& 1.05E+01 & 1.59E+01 \\
$^5I_4-^5I_7$ & 3.69E-06 & 9.00E-06 & 1.63E-05 & 2.75E-05& 6.17E-05 & 5.36E-04 & 4.19E-02 & 1.89E+00	& 6.66E+00 & 9.92E+00 \\
$^5I_4-^5I_8$ & 1.22E-11 & 1.45E-09 & 1.38E-08 & 6.30E-08& 1.18E-06 & 5.81E-05 & 1.08E-02 & 8.03E-01 & 3.01E+00 & 4.40E+00 \\
$^5I_5-^5I_6$ & 2.84E-03	& 6.88E-03 & 1.21E-02 & 1.91E-02& 3.43E-02 & 7.95E-02 & 4.32E-01 & 6.08E+00	& 1.87E+01 & 2.77E+01 \\
$^5I_5-^5I_7$ & 6.75E-05 & 1.67E-04 & 3.19E-04 & 5.95E-04& 1.34E-03 & 3.52E-03 & 9.09E-02 & 3.26E+00	& 1.12E+01 & 1.68E+01 \\
$^5I_5-^5I_8$ & 3.87E-06	& 9.41E-06 & 1.68E-05 &	2.77E-05& 5.81E-05 & 4.52E-04 &	3.92E-02 & 1.88E+00 & 6.67E+00 & 9.92E+00 \\
$^5I_6-^5I_7$ & 2.88E-03	& 6.97E-03 & 1.22E-02 & 1.94E-02& 3.49E-02 & 8.14E-02 & 4.57E-01 & 7.01E+00 & 2.21E+01 & 3.27E+01 \\
$^5I_6-^5I_8$ & 4.84E-05 & 1.19E-04 & 2.27E-04 & 4.16E-04& 9.22E-04 & 2.70E-03 & 8.48E-02 & 3.00E+00	& 1.02E+01 & 1.53E+01 \\
$^5I_7-^5I_8$ & 1.96E-03	& 4.76E-03 & 8.39E-03 & 1.34E-02& 2.44E-02 & 5.81E-02 & 3.82E-01 & 6.83E+00 & 2.18E+01 & 3.24E+01 \\
\hline\hline
\end{tabular}
\end{table*}

\subsubsection{Nd II}
\label{sec:ndIIexc}
Three excitation models were constructed and compared to assess the impact of different approximation levels, with particular emphasis on the role of RE as an intermediate approach between direct excitation and more complex treatments.
These include a direct-excitation (DE) model within the distorted-wave approximation, a RE model incorporating resonance contributions via dielectronic capture, and a benchmark model based on close-coupling R-matrix calculations.
The {\sc DARC} calculation used the same configuration set as the {\sc grasp}$^0$ calculation in table~\ref{tab:configs}, with N+1 orbitals generated for the first 6. The calculation included 24 basis functions and both parities of partial waves from a J of 0 to 49. The final effective collision strengths were then calculated across the same electron temperature range as in the {\sc autostructure} calculation. The energy levels where shifted to values from \cite{Martin_Zalubas_Hagan_1978} to ensure spectroscopic accuracy for transitions between the first 77 levels.
For the Nd {\sc ii} ion, DE was computed with {\sc autostructure} utilising the available data on NIST ASD \citep{Martin_Zalubas_Hagan_1978} to ensure the first 30 levels were shifted to be spectroscopically accurate.
The RE calculation employed the same structure and $\lambda$-values as in the DR calculation. 
Using Eq. \eqref{eq:re_eff}, the Auger-rates were post-processed to produce collision strengths for the resonant process. These are then added to the direct-component from the distorted-wave calculation to yield total strengths. This has a significant impact on forbidden and intercombination transitions, and a varied impact on the allowed transitions, where some transitions experienced a large increase at lower temperature with resonances, while others remained dominated by the direct component.  

\begin{figure*}
    \centering
    \includegraphics[width=0.8\linewidth]{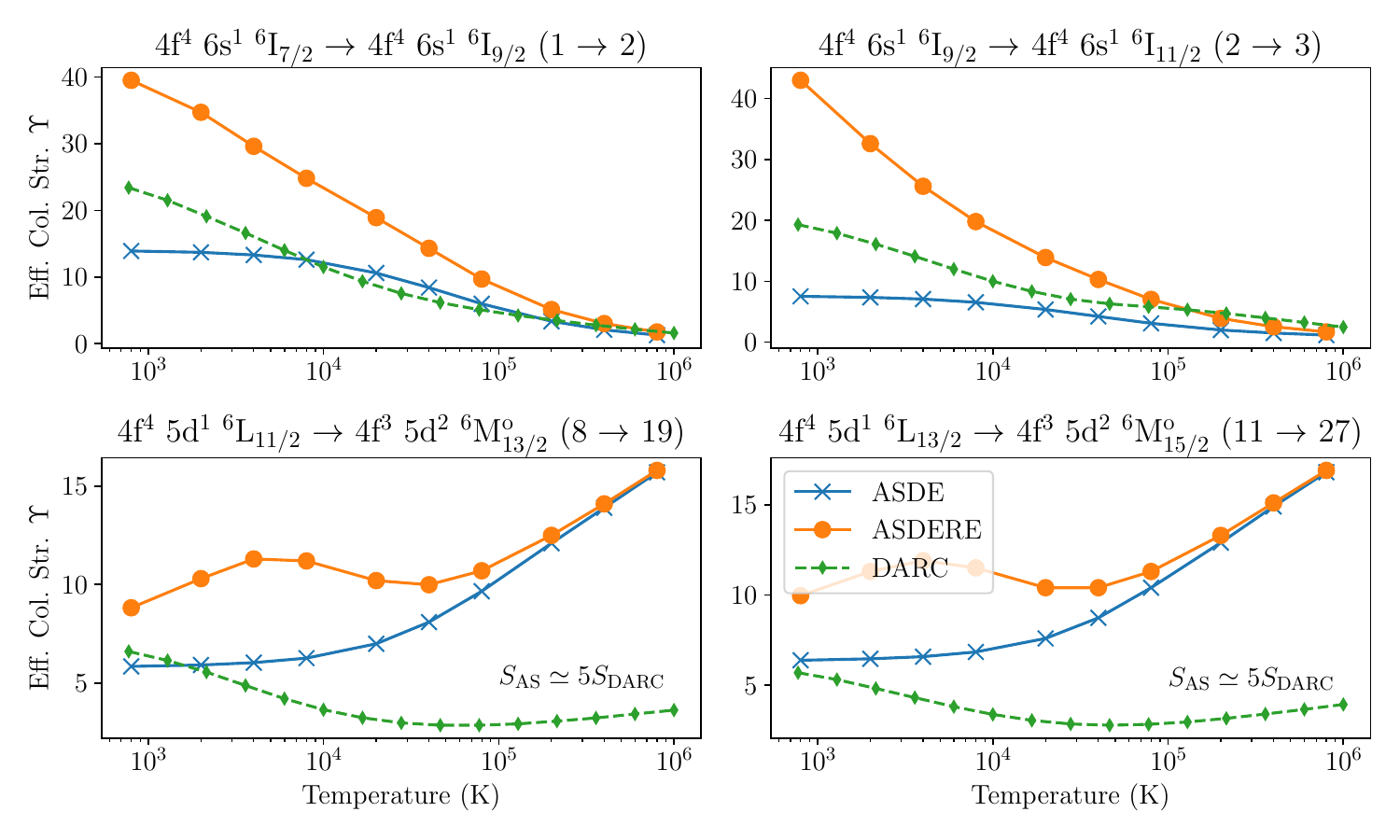}
    \caption{Values of the effective collision strengths for four Nd {\sc ii} transitions. ASDE (crosses, solid line) is the effective collision strength calculated by distorted waves. ASDERE (circles, solid line) is ASDE plus the resonant contribution from Eq. \eqref{eq:re_eff}. DARC (smaller circles, broken line) are effective collision strengths calculated by the Dirac R-matrix codes. The top two transitions are forbidden transitions within the ground term. The bottom two transitions are two strong allowed transitions within the first $\sim$ 30 levels, where the level-shifting is in agreement between all presented calculations. The two allowed lines happen to also be the two strongest allowed lines shown in Figure \ref{fig:NdII_pec}, although the emission is typically from forbidden lines. We have additionally indicated the rough difference in matrix element between AS and DARC, giving rise to the different asymptotic behaviours (see e.g \citealt{burgess1992analysis}).} 
    \label{fig:NdII_ups_res}
\end{figure*}

\begin{figure*}
    \centering
    \includegraphics[width=0.8\linewidth]{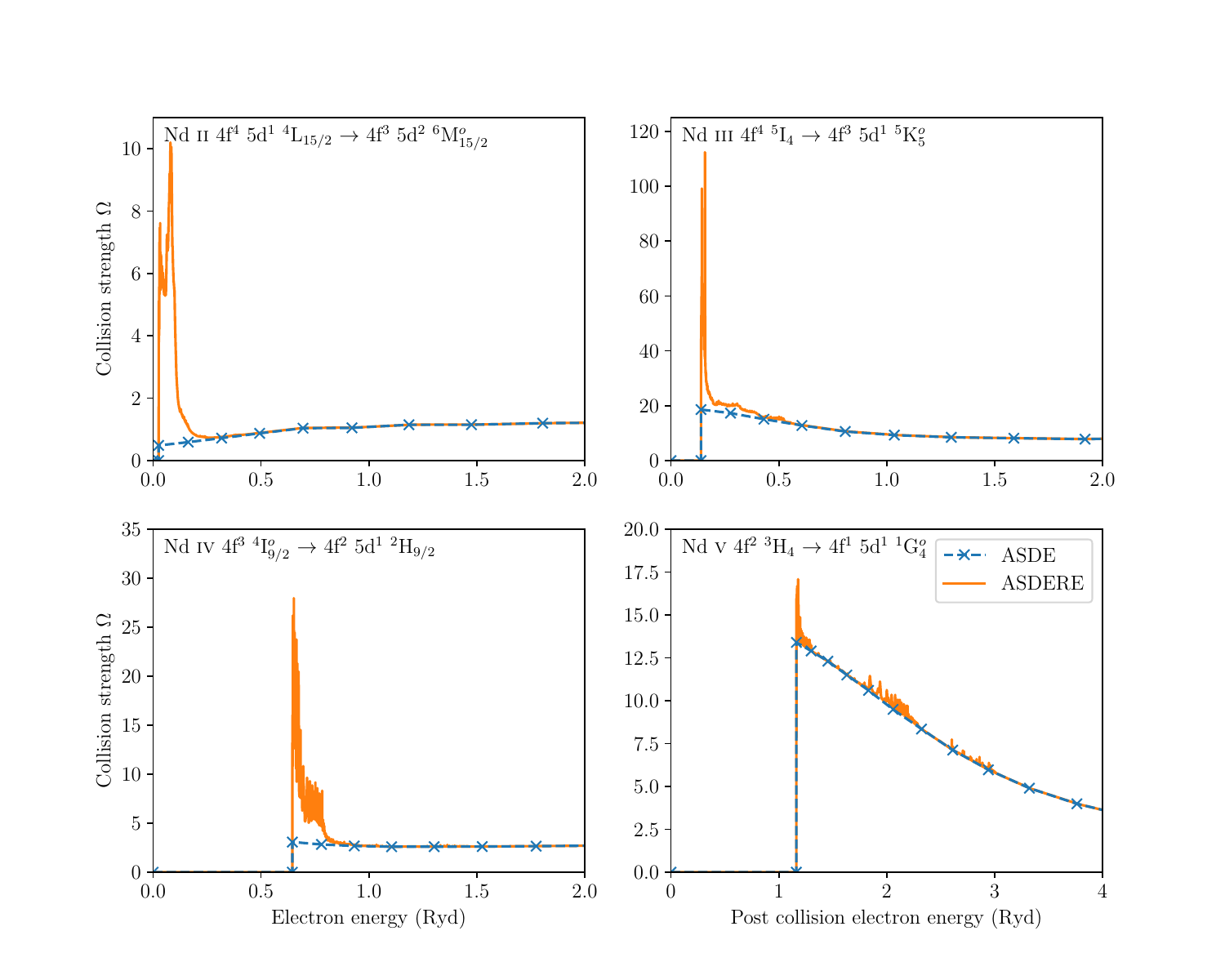}
    \caption{Values of the collision strengths for a selected dipole-allowed transition for each of Nd {\sc ii - v}. On each panel is shown the direct, and the total contribution to the collision strength.} 
    \label{fig:Ndall_omega_res}
\end{figure*}

\begin{figure*}
    \centering
    \includegraphics[width=0.8\linewidth]{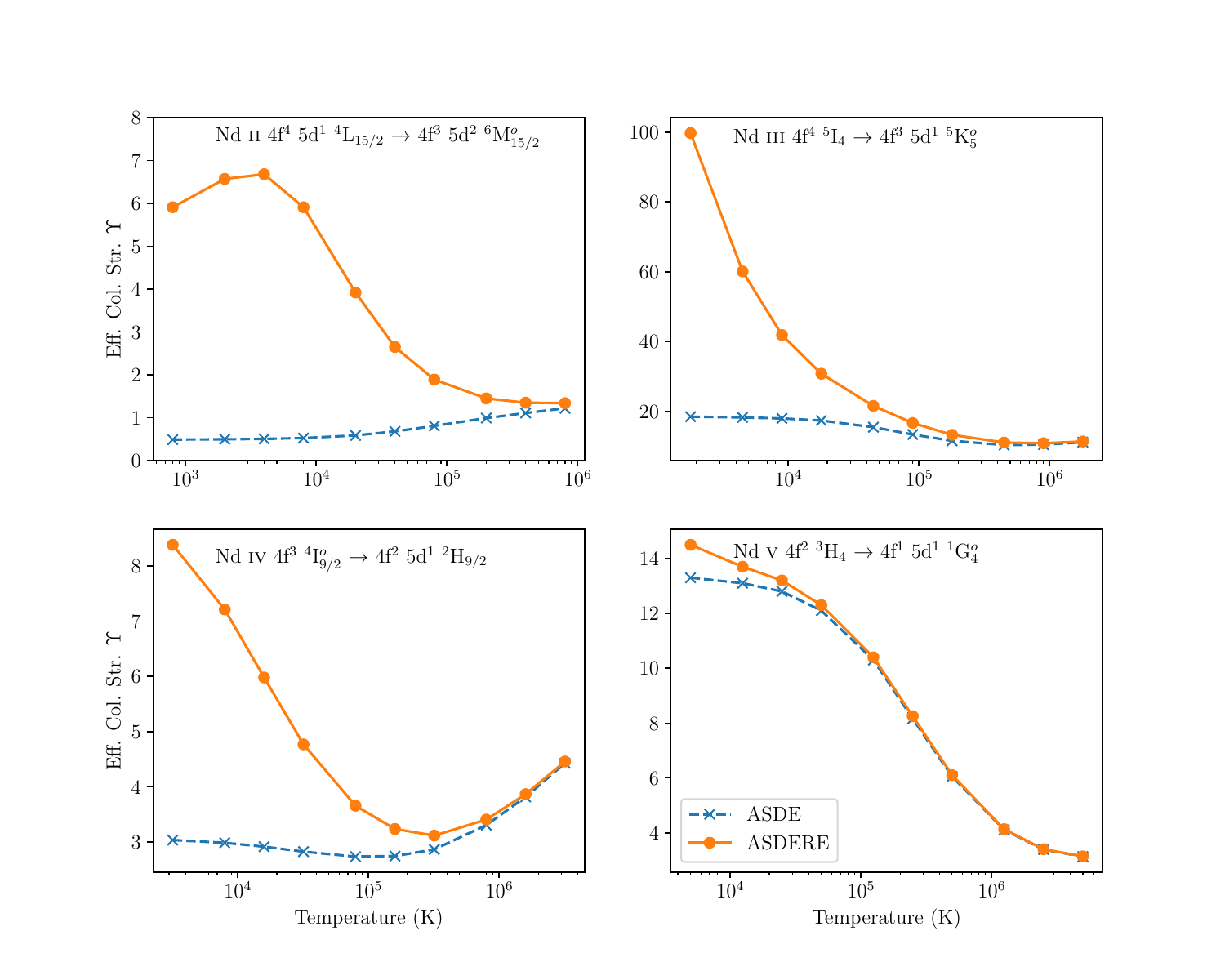}
    \caption{Same as Figure \ref{fig:Ndall_omega_res} but for the effective collision strengths.}
    \label{fig:Ndall_ups_res}
\end{figure*}

For example, the 4f$^4$ 5d$^1$ $^8$L$_{11/2}$ $\to$ 4f$^3$ 5d$^2$ $^6$M$_{13/2}^{\rm o}$ transition is shown on Figure \ref{fig:NdII_ups_res}, where the high temperature effective-collision strength is unaffected by the resonances - however the low temperatures show around a factor of two difference. This transition ($A_{j\to i} \simeq 2 \times 10^{3}$ s$^{-1}$, $\lambda \simeq 2800$ nm) is an example of an allowed MIR spectral line, characteristic of lanthanide spectra. Additionally shown on Figures \ref{fig:Ndall_omega_res} and \ref{fig:Ndall_ups_res} is the 4f$^4$ 5d$^1$ $^6$L$_{13/2}$ $\to$ 4f$^3$ 5d$^2$ $^6$M$_{15/2}^{\rm o}$ transition, which exhibits similar behaviour.  On Figure \ref{fig:NdII_ups_res}, we also show the collisions strengths as calculated by the {\sc DARC} R-matrix codes \citep{ParallelRmatrix}. While the order of magnitude is in rough agreement with the results of {\sc autostructure}, there is notably different asymptotic behaviour due to the difference in line strengths calculated by the two codes (see e.g \citealt{burgess1992analysis} for an analysis of high energy behaviour for collision strengths).

The effect of including RE is discussed in section \ref{sec:radmodel} where synthetic spectra are shown in terms of the photon-emissivity coefficient.


\subsubsection{Nd III}
A significant discrepancy is observed between the direct excitation (DE) and resonant excitation (RE) collision strengths for the dipole-allowed Nd {\sc iii} transition $4f^{4}\,{}^{5}I_{4} \rightarrow 4f^{3}5d^{1}\,{}^{5}K_{5}^{\mathrm{o}}$, as shown in Figures   \ref{fig:Ndall_omega_res} and \ref{fig:Ndall_ups_res}.
The RE values on Figure \ref{fig:Ndall_ups_res} are consistently larger than the DE results over the temperature range considered, indicating a substantial contribution from resonance enhancement, which is clearly seen from the near threshold resonance on Figure \ref{fig:Ndall_omega_res}.
This behaviour can be attributed to the complex open-$4f$ electronic structure of Nd {\sc iii}, which gives rise to a dense set of closely spaced intermediate autoionizing states near the excitation threshold. 
These resonant states provide additional excitation pathways that are not included in the smooth direct excitation background, leading to enhanced effective collision strengths in the RE calculation. 
The effect is particularly pronounced at low and intermediate temperatures, where near-threshold resonances make the largest contribution to the Maxwellian-averaged collision strengths.

\subsubsection{Nd IV}
The collision strengths for a selected dipole-allowed transition for Nd {\sc iv} are shown on Figures \ref{fig:Ndall_omega_res} and \ref{fig:Ndall_ups_res}.
The difference between the collision strengths from the direct and resonant excitation models is most pronounced at low energies and temperatures. This reflects the strong influence of near-threshold resonances, which significantly enhance the effective collision strengths through resonant excitation pathways. 
As a result, the inclusion of resonant contributions leads to noticeable increases in the cooling function in this temperature regime, highlighting the importance of accurately capturing these processes for reliable plasma modelling.

\subsubsection{Nd V}
As with all previous ion stages, DE and RE was computed for Nd {\sc v}. From the plotted collision strengths in figures \ref{fig:Ndall_omega_res} and \ref{fig:Ndall_ups_res}, the resonant contribution appears to be very minor for this ion. 
The effect of resonant contribution on collisional radiative models is discussed in section \ref{sec:radmodel}.

\section{Benchmarking Excitation with Collisional Radiative Modelling}\label{sec:radmodel}

\begin{figure*}
    \centering
    \includegraphics[width=\linewidth]{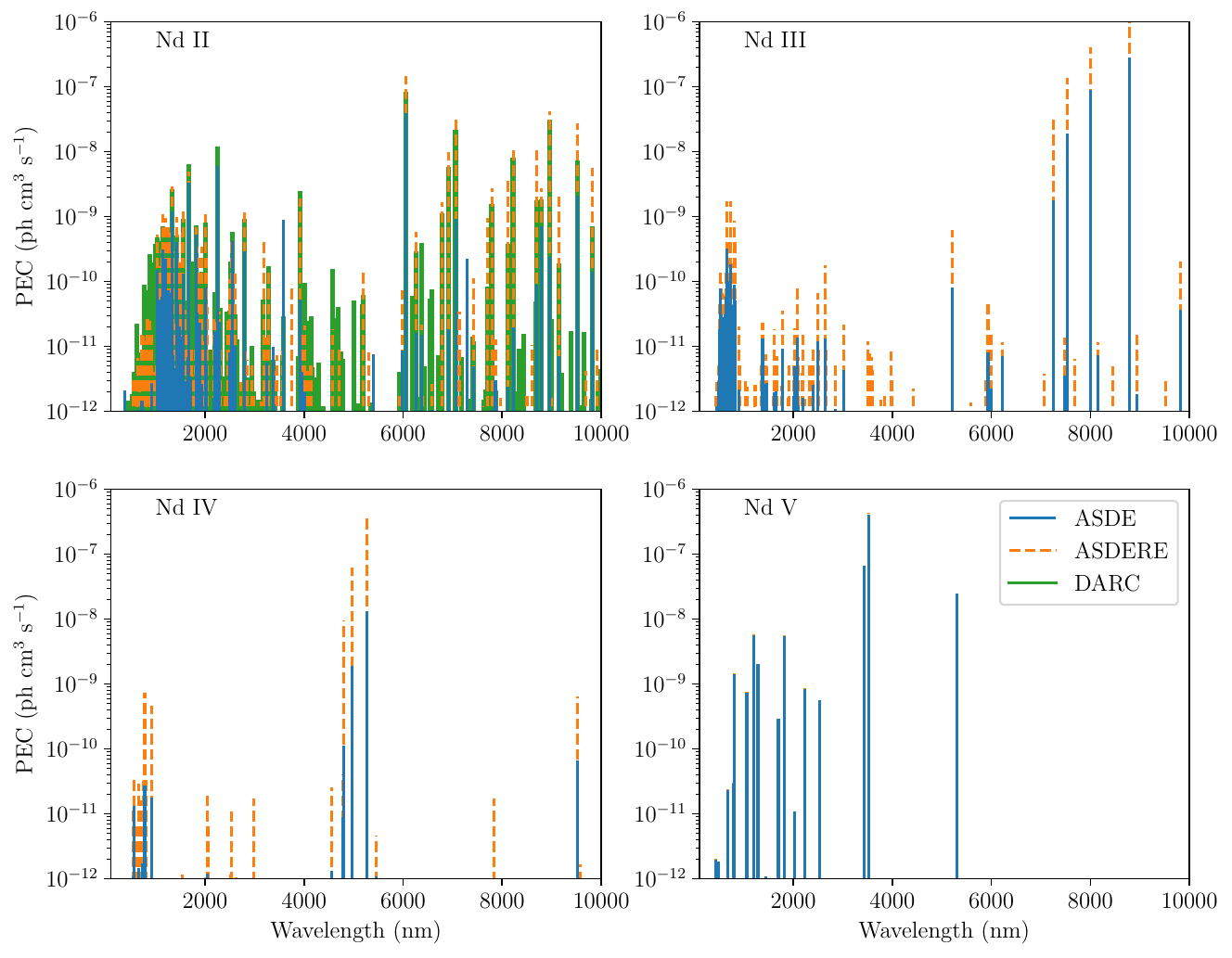}
    \caption{Photon Emissivity Coefficient (PEC) stick spectra for Nd II–V as a function of wavelength, at an electron density of $n_e = 10^4$ cm$^{-3}$ and temperature $T_e=2900$K.} Nd II includes comparisons between DE, RE, and DARC calculations, while Nd III–V show DE and RE results. Resonant excitation increases the PECs for the lower ion stages, particularly Nd II–IV, whereas for Nd V the DE and RE spectra overlap closely, indicating negligible resonant enhancement.
    \label{fig:NdII_pec}
\end{figure*}

Using the radiative parameters and collision strengths, we are able to solve the ion level population on an ion-by-ion basis. Calculations involving a charge balance are reserved for future work employing a detailed description of non-thermal electron ionization.
The steady state of the rate equation,
\begin{equation}
	    \frac{\dd N_i}{\dd t} = \sum_{j} C_{ij} N_j,\label{eq:cr}
\end{equation}
is found using the {\sc colradpy} code \citep{Johnson_Loch_Ennis_2019}. Where $N_i$ and $N_j$ are the population of levels $i$ and $j$, $C_{ij}$ is the sum of the rates from $j$ into $i$ and if $i$ and $j$ are the same then $C_{ij}$ would be the rates out of that level. Studies have shown that the steady-state solution of the collisional-radiative equation is generally adequate for kilonova modelling \citep{pognan_validity_2022}. A notion of the emission-per-atom at unit density of a particular transition is given by the Photon-Emissivity-Coefficient, 
\begin{equation}
    \text{PEC}_{j \to k} = A_{j \to k} f_j /n_e ,
\end{equation}
where $A_{j \to k}$ is the transition rate from $j$ to $k$, $n_e$ is the electron density and the population $f_j$ is the fraction of the ion in level $j$, i.e the populations normalized such that $\sum_j f_j=1$.
This quantity is shown on Figure \ref{fig:NdII_pec} for each of the four systems considered in this work, at a density of $10^4$ cm$^{-3}$ and electron temperature of $0.25$ eV $\approx$ 2900K. We compare between all data sets presented in this work. It can be seen that the resonant contribution is perhaps most evident in the infrared - which is mostly dominated by low-lying forbidden transitions. In the Nd {\sc ii} case, it is seen that there is generally reasonable agreement, but some notable discrepancy between the direct PEC and that of {\sc darc}. It is however generally seen that the inclusion of resonances with the direct distorted wave brings the PEC in line with, or in some cases in slight excess of the {\sc darc} calculation. This is likely due to the generally higher collision strengths as seen in Figure \ref{fig:NdII_ups_res}. None the less, it is clear that the lower computational cost of the isolated resonance approximation gives a good representation of the plasma dynamics when compared to the R-matrix collision strengths. This is further emphasised later in this section in the discussion of cooling. It is additionally notable that there spectral lines from either calculation (e.g around 5000 nm) that do not coincide with the other. This is due to differences in atomic structure and wavelength calibration between the two calculations. The ions Nd {\sc iii - v} show familiar trends with generally higher emission intensity when resonances are included due to enhanced collision strengths.

From this, we are able to calculate the total radiative power loss (PLT) per unit electron density,
\begin{equation}
    \text{PLT} = \sum_j\sum_{k<j} E_{kj} \text{PEC}_{j \to k},
\end{equation}
where $E_{ij}$ is the energy of the transition between $i$ and $j$.
This is related to the cooling function  $\Lambda_i$ of ion stage $i$ as described in \cite{hotokezaka_nebular_2021} \footnote{ The normalized cooling function from \citet{hotokezaka_nebular_2021} is,
\begin{align*}
    \frac{\Lambda_i}{n_en_i}   &=  \sum_{j} \sum_{k<j} \frac{n_{i,j}n_e}{n_i}E_{kj}A_{kj} \beta_{kj}, \\
    &= \sum_{j} \sum_{k<j}E_{kj}\text{PEC}_{j \to k},\\
    &=\text{PLT},
\end{align*} where $n_i$ is the total number density of ion stage $i$ and $n_{i,j}$ that of level $j$. Thus $f_j = n_{i,j}/n_i$.  }

\begin{figure*}
    \centering
    \includegraphics[width=\linewidth]{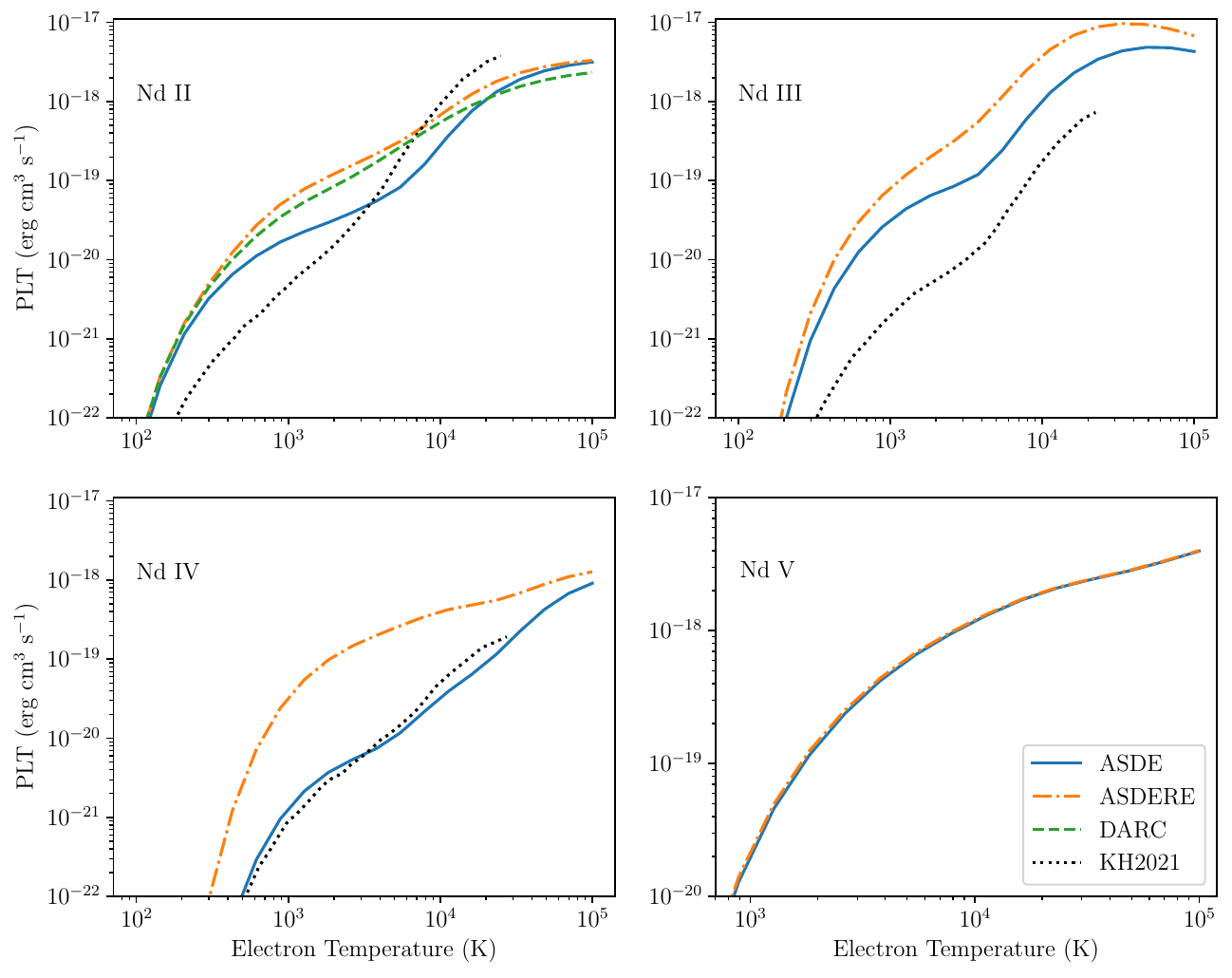}
    \caption{Cooling functions produced using the atomic data outlined within this work, compared where possible with the results of \citet{hotokezaka_nebular_2021}. Here, {\sc asde} denotes a cooling function computed with only DE considered; {\sc asdere} includes contributions from both direct and resonant excitation; and {\sc darc} includes both processes in a close-coupling formalism using the Dirac $R$-matrix method. }
    \label{fig:NdCooling}
\end{figure*}

We calculate the PLT for Nd {\sc ii - iv} using each of the available datasets and compare with \cite{hotokezaka_nebular_2021}. They employed the approximation of \cite{van1962rate} for allowed lines, and a constant value of unity for forbidden lines. This constant value was calculated based on forbidden collision strengths from the {\sc hullac} code \citep{bar2001hullac}. We show this comparison on Figure \ref{fig:NdCooling}.
It is seen that the cooling from the direct-excitation-only calculation agrees generally well with that of \cite{hotokezaka_nebular_2021}, although the dependence of the cooling on both the excitation and radiative decay rates make direct comparison difficult. Generally there is agreement within an order of magnitude. The calculations indicate that the inclusion of resonances increases the cooling by a significant amount. This is perhaps unsurprising, as larger collision strengths will push the collisional-radiative matrix towards the limiting case of Boltzmann populations. The similarity is most obvious in the Nd {\sc iv} case, where the cooling calculated with DE rates only closely matches that of \cite{hotokezaka_nebular_2021}. Notably, the resonant contribution is most obvious in this case with a $\sim 2$ dex difference in the cooling profile.

No significant difference is observed between the cooling functions derived from the DE and RE models for Nd {\sc v}. This indicates that resonance contributions play a relatively minor role for this ion.
This behaviour is consistent with the expectation that, for higher ionization stages, the density of near-threshold resonances is reduced and resonances appear at higher energies, reducing their contributions at the temperatures considered.
Even if small differences exist within individual transitions, the cooling functions are integrated quantities and so small resonance enhancements will be averaged out.
As a result, the excitation is dominated by the direct process and including the RE does not significantly alter the resulting cooling rates.

\section{Conclusions}
In this work, we have presented comprehensive recombination and electron-impact excitation data for the first four ionization stages of neodymium. 
The atomic structures of all ions were computed in a systematic and consistent manner using the internal optimization scheme of {\sc autostructure}.
The temperatures involved within the kilonova environment enabled focus to be on the low-temperature atomic data.
This involved focusing precision on the low-lying resonance positions and neglecting precision at higher temperatures.
Emphasis should be made that the atomic data produced using this tactic should be used with caution for high-temperature applications as the data has not been optimized for this application.
For the ions considered here, DR provides the dominant contribution to the total recombination rate coefficients over much of the temperature range. 
The notable exception is Nd {\sc ii}, where RR dominates at low temperatures, with DR becoming increasingly important as the temperature rises.

Three electron-impact excitation models were constructed for Nd {\sc ii}. 
Radiative and direct excitation data were computed using {\sc autostructure}, while a full close-coupling calculation employing the GRASP/R-matrix framework was used to produce a benchmark (“gold-standard”) excitation model. 
These models were compared to assess their relative accuracy and computational cost, providing justification for adopting the RE approach as a practical intermediate solution when full R-matrix calculations are computationally demanding.
The RE model captures excitation mediated by resonant processes, arising from dielectronic capture into autoionizing states followed by radiative stabilization, which are entirely absent from the purely non-resonant DE model.
The use of the RE method provides a computationally efficient means of incorporating key resonance effects, making it a practical precursor to more rigorous R-matrix calculations when detailed treatment of specific ions is required.

These results provide a consistent set of atomic data for Nd {\sc i - v} and demonstrate their relevance for plasma modelling applications, particularly in environments where accurate treatment of excitation and recombination processes is required.
The approach presented here can be directly applied to other heavy species, offering a consistent framework for extending atomic datasets to similarly complex ions.

\section*{Acknowledgements}
N.F acknowledges the Science and Technology Facilities Council (STFC) part of the UK Research and Innovation body for their support via studentship, grant number 2931114.
This research used resources of the National Energy Research Scientific Computing Center (NERSC), a Department of Energy Office of Science User Facility using NERSC award FESERCAP0031596.

L.P.M, M.M and C.P.B are funded/Co-funded by the European Union (ERC, HEAVYMETAL, 101071865). Views and opinions expressed are however those of the author(s) only and do not necessarily reflect those of the European Union or the European Research Council. Neither the European Union nor the granting authority can be held responsible for them. 

This work used the DiRAC Data Intensive service (DIaL3) at the University of Leicester, managed by the University of Leicester Research Computing Service on behalf of the STFC DiRAC HPC Facility (www.dirac.ac.uk). The DiRAC service at Leicester was funded by BEIS, UKRI and STFC capital funding and STFC operations grants. DiRAC is part of the UKRI Digital Research Infrastructure.

This work used the DiRAC Extreme Scaling service (Tursa) at the University of Edinburgh, managed by the EPCC on behalf of the STFC DiRAC HPC Facility (www.dirac.ac.uk). The DiRAC service at Edinburgh was funded by BEIS, UKRI and STFC capital funding and STFC operations grants. DiRAC is part of the UKRI Digital Research Infrastructure.

\section*{Data Availability}
The data supporting this article will be made publicly available in the Zenodo repository at \url{https://doi.org/10.5281/zenodo.20267654} .
The datasets will also be available at \url{https://open.adas.ac.uk} in due course.



\bibliographystyle{mnras}
\bibliography{references1}



\appendix


\bsp	
\label{lastpage}
\end{document}